\documentclass[aps,prd,floatfix,10pt,nofootinbib,showpacs,notitlepage,longbibliography]{revtex4-1} 
\usepackage{amsmath}
\usepackage{amsfonts}
\usepackage[colorlinks=true,linkcolor=blue,citecolor=blue,urlcolor=blue]{hyperref}
\usepackage{amssymb}
\usepackage{graphicx}
\usepackage{enumerate}
\usepackage{csquotes}

\newcommand{\nn}{\nonumber \\}
\newcommand{\eq}[1]{Eq.~(\ref{#1})}
\newcommand{\fig}[1]{Fig.~\ref{#1}}

\newcommand{\bsub}{\begin{subequations}}
\newcommand{\esub}{\end{subequations}}
\newcommand{\be}{\begin{eqnarray}}
\newcommand{\ee}{\end{eqnarray}}
\newcommand{\om}{\ensuremath{\omega}}
\newcommand{\omo}{\ensuremath{{\lambda}}}
\newcommand{\omi}{\ensuremath{{\omega}}}

\newcommand{\X}{\ensuremath{X}} 
\newcommand{\psiu}[1]{\ensuremath{\psi_#1^{u}}}
\newcommand{\psiv}[1]{\ensuremath{\psi_#1^{v}}}
\newcommand{\phiu}[1]{\ensuremath{\phi_#1^{u}}}

\newcommand{\psil}[1]{\ensuremath{\lp \psi_{#1}^{u, \leftarrow}} \rp^*}

\newcommand{\psir}[1]{\ensuremath{\lp \psi_{#1}^{u, \rightarrow}}\rp^*}

\usepackage{float}

\newcommand{\pd}{\ensuremath{\partial}}

\newcommand{\lp}{\ensuremath{\left(}}
\newcommand{\rp}{\ensuremath{\right)}}
\newcommand{\bi} {\begin{itemize}}
\newcommand{\ei} {\end{itemize}}
\newcommand{\ben}{\begin{enumerate}}
\newcommand{\een}{\end{enumerate}}
\newcommand{\bmat}{\begin{pmatrix}}
\newcommand{\emat}{\end{pmatrix}}

\begin{abstract}
In Ho\v{r}ava and Einstein-\ae ther theories of modified gravity, in spite of the violation of Lorentz invariance, spherically symmetric stationary black hole solutions possess an inner {\it universal} horizon which separates field configurations into two disconnected classes. We compute the late time radiation emitted by a dispersive field propagating in such backgrounds. We fix the initial conditions on stationary modes by considering a regular collapsing geometry, and imposing that the state inside the infalling shell is vacuum. We find that the mode pasting across the shell is adiabatic at late time (large inside frequencies). This implies that large black holes emit a thermal flux with a temperature fixed by the surface gravity of the Killing horizon. In turn, this suggests that the universal horizon should play no role in the thermodynamical properties of these black holes. 
\end{abstract}

\begin{document}

\title{Black hole radiation in the presence of a universal horizon}


\author{Florent Michel}\email[]{florent.michel@th.u-psud.fr} 
\author{Renaud Parentani}\email[]{renaud.parentani@th.u-psud.fr}
\affiliation{Laboratoire de Physique Th\'eorique, CNRS UMR 8627, B{\^{a}}timent\ 210,
 \\Universit\'e Paris-Sud 11, 91405 Orsay CEDEX, France}

\maketitle

\section{Introduction}
 
The laws of black hole thermodynamics are firmly established in Lorentz invariant theories, and they play a crucial role in our understanding of black hole physics~\cite{wald1994quantum}. In particular, the entropy and the temperature are governed by the area and surface gravity of the event horizon. In Lorentz violating theories, the status of these laws is unclear because essential aspects are no longer present~\cite{Jacobson:2001kz, Dubovsky:2006vk, Jacobson:2008yc, Betschart:2008yi, Blas:2011ni, Busch:2012ne}. For instance, the thermality of the Hawking flux is inevitably lost in the presence of high frequency dispersion, although it is approximatively recovered for large black holes, i.e., when the surface gravity $\kappa$ is much smaller than the UV scale $\Lambda$ setting the high frequency dispersion~\cite{Macher:2009tw}. 

The origin of the difficulties can be traced to the fact that the event horizon no longer separates the outgoing field configurations into two disconnected classes. In fact, when the dispersion is superluminal, it can be crossed by outgoing radiation. However, it was recently discovered that in some theories of modified gravity such as Ho\v{r}ava gravity~\cite{Horava:2009uw, Sotiriou:2010wn, Janiszewski:2014iaa} and Einstein-\ae ther~\cite{Jacobson:2000xp, Eling:2004dk, Eling:2006ec, Barausse:2011pu}, spherically symmetric
black hole solutions possess a second inner horizon. This horizon, named {\it universal}, cannot be crossed by outgoing configurations, even for superluminal dispersion relations which allow for arbitrarily large group velocities. (The difficulty mentioned in~\cite{Jacobson:2001kz} is thus evaded.) Following this discovery, it has been argued that the universal horizon should play a key role in the thermodynamics of such black holes. Even though they seem to obey a first law~\cite{Berglund:2012bu, Horavasummary}, a key question concerns the temperature of the Hawking radiation they emit. Would it be essentially governed by the (higher) surface gravity of the universal horizon, or would it still be fixed by the surface gravity $\kappa$ of the Killing horizon?  

Two recent works concluded that the universal horizon emits a steady radiation with properties governed by its surface gravity. Because of the complicated nature of the field propagation near that horizon, this conclusion was indirectly obtained, in~\cite{Berglund:2012fk}, by making use of a ``tunneling method'', and, in~\cite{Cropp:2013sea}, by analyzing the characteristics of the radiation field. In the present paper, we reexamine this question by performing a direct calculation and reach 
the opposite conclusion that no radiation is emitted from the universal horizon at late time.

We proceed as follows. As in the original derivation of Hawking~\cite{Hawking:1974sw}, we identify the boundary conditions on the outgoing modes in the near vicinity of the universal horizon by considering a simple collapsing shell geometry, and by assuming that the state of the field is vacuum inside. We then compute the mode mixing across the shell between inside modes propagating outwards $\phi_\omi^{\rm in}$, and outside stationary modes with a fixed Killing frequency $\psi_\omo$. The late time behavior is obtained by sending the inside frequency $\omi \to \infty$. In this limit, we show that the scattering coefficients involving modes with opposite norms vanish. This result can be understood from the fact that the modes $\psi_\omo$ are accurately described by their WKB approximation in the immediate vicinity of the universal horizon. In other words, the pasting across the shell is adiabatic in the limit $\omi \to \infty$. Hence, for large outgoing radial momenta, the state of the field outside the shell is the usual vacuum, as explained in~\cite{Brout:1995wp}.

It then remains to propagate these high momentum dispersive modes from the universal horizon till spatial infinity. This propagation has already been studied in detail; see~\cite{Coutant:2011in} for a recent update. It establishes that large black holes emit a stationary flux which is (nearly) thermal, and with a temperature approximatively given by the standard relativistic value. In other words, the robustness of the Hawking process, i.e. its insensitivity to high frequency dispersion which was first established in~\cite{Unruh:1994je}, is now extended to black holes with a universal horizon.

From this it is tempting to conclude that the laws of black hole thermodynamics should also be robust, and they should involve the properties of the Killing horizon. This conclusion is reinforced by the fact that the field configurations propagating on either side of a universal horizon come from two disconnected Cauchy surfaces, and are highly blueshifted. Hence, it seems that no Hadamard condition of regularity~\cite{Busch:2012ne} could be satisfied on the universal horizon. This raises the question of the fate of the universal horizon; see~\cite{Blas:2011ni}. This difficult question shall not be discussed in the present work.

Appendix~\ref{app:details} gives the details of the calculation which is summarized in the main text. In Appendix~\ref{app:acc} we compare our model with previously-studied dispersive ones without a universal horizon, and show the role of the acceleration of the preferred frame. Appendix~\ref{app:Hawking} shows the results of numerical simulations confirming the approximately thermal character of the emission at infinity governed by the surface gravity of the Killing horizon. 

\section{Massless relativistic scalar field in a collapsing shell geometry}

In this section, we briefly review the computation of the Hawking radiation emitted at late time in a collapsing geometry~\cite{Hawking:1974sw}. Although these concepts are well known, we present them in a way which prepares the more involved calculation of the late time flux when dealing with a dispersive field in the presence of a universal horizon. As explained in the Introduction, we shall use a direct calculation which consists of pasting the modes across the infalling shell. We closely follow the derivation of~\cite{Massar:1997en}.

\begin{figure}[h!]
\begin{center}
\includegraphics[width=0.5\linewidth]{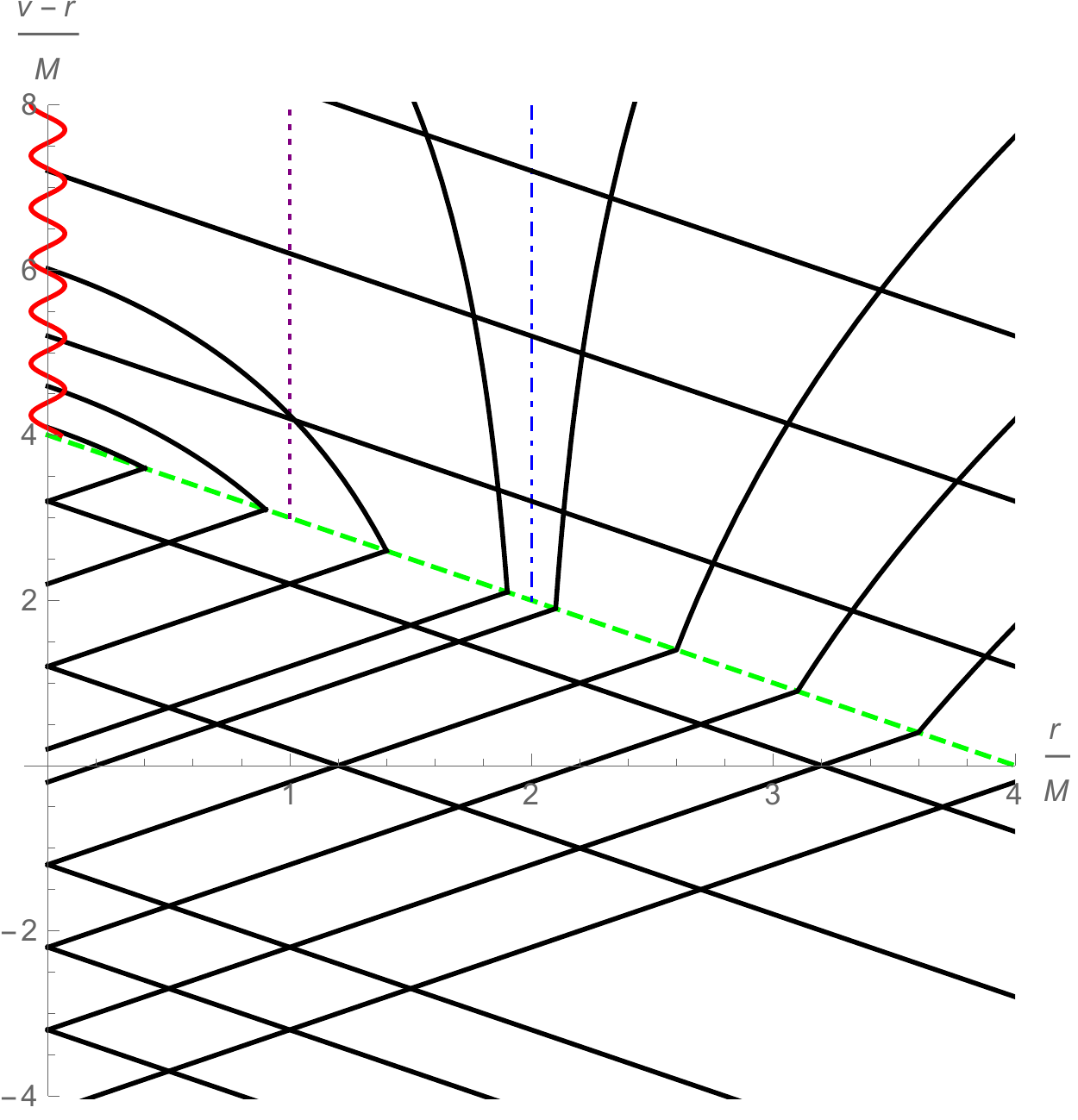} 
\includegraphics[width=0.35\linewidth]{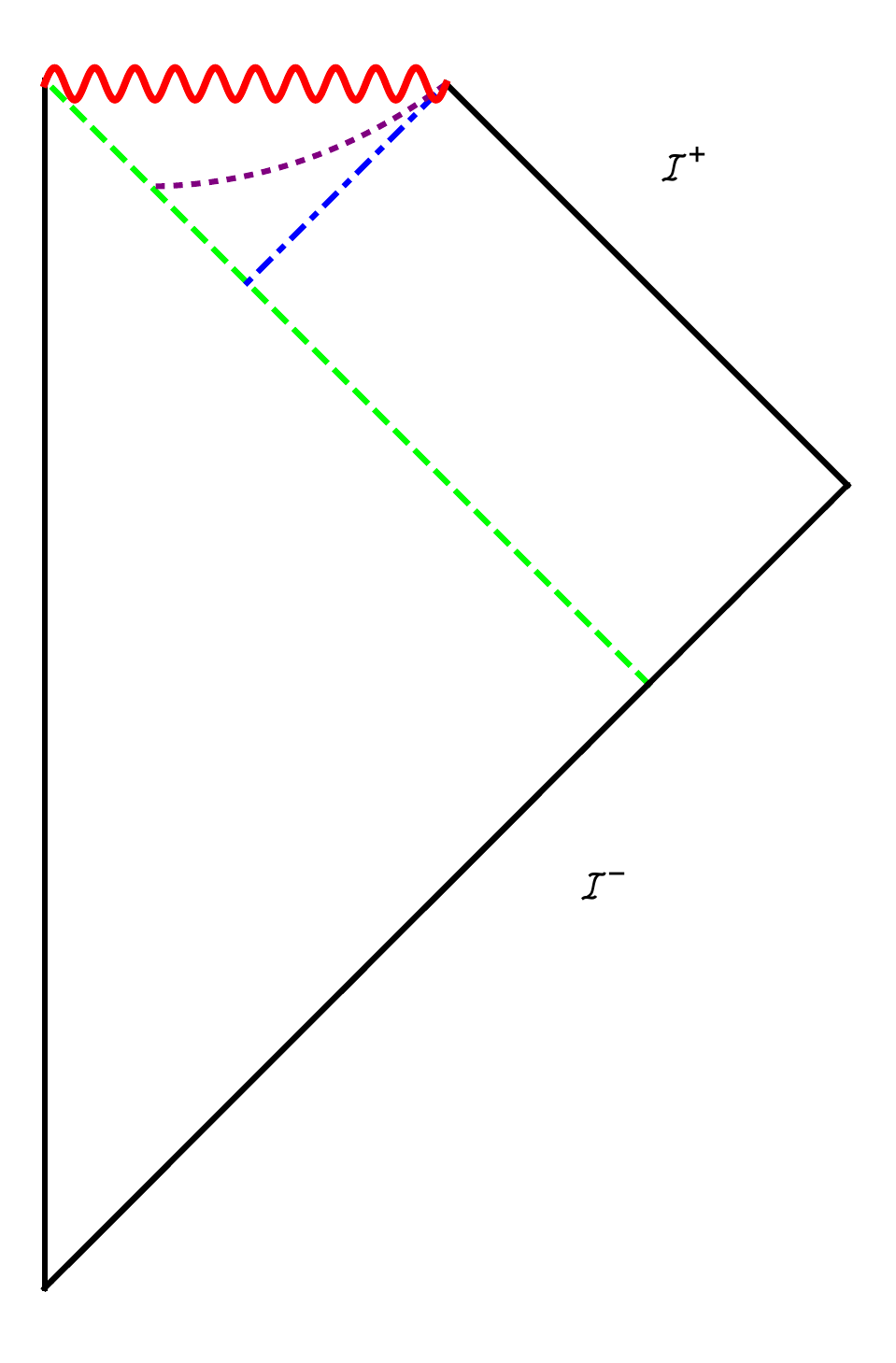}
\end{center}
\caption{(Left panel) Null radial geodesics in the $(v-r,r)$ plane in units of $M$. The coordinate $v-r$ coincides with the Minkowski time $T$ inside the mass shell, and with the Schwarzschild time $t_S$ for $r \to \infty$. The solid black lines are null radial geodesics which are reflected on $r=0$. The dashed green line represents the trajectory of the null shell $v = 4M$. The blue, dot-dashed one shows the Killing horizon at $r=2M$ outside the shell. The dotted purple line shows the locus $r=M$, $v>4M$ which will play a crucial role in Section~\ref{sec:Horava}. The wavy line shows the singularity, located at $r=0, \, v>4M$. (Right panel) Penrose-Carter diagram of the collapsing shell geometry. The vertical line corresponds to $r=0, \, v<4M$. $\mathcal{I}^-$ corresponds to $u,U \to -\infty$, and $\mathcal{I}^+$ to $v \to \infty$. 
} \label{fig:RelST} 
\end{figure}

For simplicity, we consider an infalling spherically symmetric lightlike thin shell. In this case, it is particularly appropriate to work with advanced Eddington-Finkelstein (EF) coordinates $v,r$, where $v$ is the advanced null time. At fixed $r$, one has $dv/dt_S = 1$, where $t_S$ is the usual Schwarzschild time. Hence, outside the shell, the stationary Killing field $K^\mu \partial_\mu$ is simply $\pd_v$. On both sides of the shell taken to be $v=4M$, the line element reads
\be \label{eq:met}
ds^2 = \lp 1-\frac{2M(v)}{r} \rp dv^2 - 2 dv dr-r^2 \lp d \theta^2 + \sin(\theta)^2 d \varphi^2 \rp. 
\ee 
where $M(v) = \Theta(v - 4M) M $. These coordinates cover the entire space-time, shown in the right panel of \fig{fig:RelST}. On the left panel, the infalling and outgoing null radial geodesics are represented in the $(v-r,r)$ plane. One clearly sees that the Killing horizon (where the norm $K^\mu K_\mu$ vanishes) divides the outgoing geodesics into two separate classes. We work in Planck units: $c = \hbar = G = 1$. 

Let $\Phi$ be a massless real scalar field with the action
\be \label{eq:relativisticaction}
S = \int d^4 x \sqrt{-g} \lp \pd_\mu \Phi \rp \lp \pd^\mu \Phi \rp. 
\ee
We define $\psi \equiv r \Phi$ and consider radial solutions independent of $\lp \theta, \varphi \rp$. Inside the shell, for $v < 4M$, we introduce the null outgoing (affine) coordinate $U = v-2r $. Outside the shell and for $r > 2M$,  we introduce the null coordinate
\be
u = v-2r_K^* \in (-\infty, \infty), 
\ee
where $r^*_K = r + 2 M \ln \left\lvert {r}/{2M} - 1 \right\rvert$ is the usual tortoise coordinate, which diverges on the Killing horizon. To cover the region inside the Killing horizon, one needs another coordinate $u_L = - (v - 2r_K^*)$.~\footnote{This sign guarantees that $dU/du_L$ is positive. As we shall see in Section~\ref{sec:Horava}, a similar sign must be taken when studying a dispersive field on both sides of a universal horizon.} The field equation then reads
\be 
\left\lbrace 
\begin{array}{cc}
\pd_U \pd_v \psi = 0,  & v<4M , \\
\lp \pd_u \pd_v + \lp 1- \frac{2M}{r} \rp \frac{2M}{r^3} \rp \psi = 0 , & v>4M .
\end{array}
\right. 
\ee
For simplicity, we neglect the potential engendering the grey body factor and work with the conformally invariant equations $\pd_U \pd_v \psi = \pd_u \pd_v \psi = 0$. The solutions can be decomposed as
\be 
\psi(u,v) = \psi^{u}(u) + \psi^v(v),
\ee
and similarly for $v<4M$ with $u$ replaced by $U$. The infalling $v$ sector and the outgoing $u$ sector completely decouple. Moreover, the $v$ modes $\psi^v$ are regular across the horizon and play no role in the Hawking effect. We thus consider only the $u$ modes, and, to lighten the notations, we no longer write the upper index $u$ on outgoing modes. 

To compute the global solutions, we need the matching conditions across the null shell. In the present case, $\psi$ is continuous along $v=4M$. Hence $\psi_{\rm inside}(U) = \psi_{\rm outside}(u(U))$, where the relation between null coordinates is
\be   
u(U) = U - 4 M \ln \lp \frac{ - U }{4M} \rp  ,
\label{uU}
\ee
for $r > 2M$ ($U < 0$). For $r < 2M$ ($U > 0$), one has $u_L(U) =  - u(|U|)$.

To obtain the Hawking flux one needs to relate the $in$ modes $\phi_\omi^{\rm in}$ characterizing the vacuum inside the shell, to the $out$ modes $\phi_\omo^{\rm out}$ characterizing the asymptotic outgoing quanta with Killing frequency $\omo$. In the internal region, a complete orthonormal basis of positive-norm modes is provided by the plane waves
\be 
\phi_\omi^{\rm in} \equiv \frac{e^{-i \omi U}}{2 \sqrt{\pi \omi}},
\label{relat-in-mode}
\ee
where $\omi \in \mathbb{R}^+$ is the inside frequency $i \partial_U$. In the external region, the (positive-norm) stationary modes for $r > 2M$, are
\be  
\phi_\omo^{\rm out} 
\equiv \Theta(r-2M)\, \frac{e^{-i \omo u}}{2 \sqrt{\pi \omo}} , \; \omo \in \mathbb{R}^+.
\label{relat-out-mode}
\ee
A similar equation defines $\phi_\omo^{(L)}(u_L)$ in the trapped region, for $r< 2M$. The modes $\phi_\omo^{\rm out} , \phi_{\omo'}^{(L)}$ and their complex conjugate form a complete orthonormal basis. One easily verifies that the conserved scalar product for the $u$ modes can be written as 
\be 
\lp \psi_1, \psi_2 \rp = i \int^\infty_{- \infty} du \lp \psi_1^* \pd_u \psi_2 - \psi_2 \pd_u \psi_1^* \rp. 
\ee

The Bogoliubov coefficients encoding the Hawking flux are then given by the overlaps between the two sets of modes:
\begin{align} 
\alpha_{\omo, \omi} &= \lp \phi_{\omo}^{\rm out}, \phi_{\omi}^{\rm in}  \rp, \nn 
\beta_{\omo, \omi} &= \lp (\phi_{\omo}^{\rm out} )^{*}, \phi_{\omi}^{\rm in}  \rp.
\label{BogC}
\end{align}
Using $u(U)$ of \eq{uU}, they can be computed explicitly; see~\cite{Massar:1997en} for details. The late time behavior is obtained by sending the inside frequency ${\omi \to \infty}$. In this limit, one recovers the standard thermal result
\be 
\left\lvert \frac{\beta_{\omo, \omi}}{\alpha_{\omo, \omi}} \right\rvert^2 \mathop{\sim}_{\omi \to \infty} e^{-8 \pi M \omo} .
\label{betovera}
\ee

To prepare for the forthcoming analysis, it is instructive to compute the Bogoliubov coefficients by the saddle point method~\cite{Parentani:1992me,Brout:1995rd}. For the $\alpha_{\omo, \omi}$ coefficient, when $\omi \gg \omo$, i.e., at late time, the location of the saddle is given by
\be
\omo = \omi \, 
\lp e^{-\kappa (u - u_0)} + O(e^{- 2 \kappa (u- u_0)})\rp 
,  
\label{CarterL}
\ee
where $u_0$ is a constant which drops out of the late time flux. (In the present model, $u_0$ vanishes.) From this equation we recover the time-dependent redshift relating $\omi$, the large frequency emitted from the collapsing star, to $\lambda$, the frequency received at infinity and measured using the proper time of an observer at rest. In particular, we recover the characteristic exponential law governed by the surface gravity $\kappa = \pd_u \ln U(u) =  \frac{1}{4M}$. Had we considered a collapsing shell following a (regular) infalling timelike curve, \eq{CarterL} would still have been obtained at late $u - u_0$ time. 

This is the kinematical root of the universality of Hawking radiation in relativistic theories. Indeed, when studying the $\beta_{\omo, \omi}$ coefficient, one finds that the saddle point is now located at $\omo = - \omi e^{-\kappa u_{\rm s.p.}}$. When taking into account the fact that the integration contour should be deformed in the lower $u$-complex plane, one finds that $u_{\rm s.p.}$ has an imaginary part $\Im \lp u_{\rm s.p.} \rp = - \pi / \kappa$, whereas its real part in unchanged. This gives a relative factor $\exp \lp  - \pi \omo / \kappa \rp$ with respect to the $\alpha_{\omo, \omi}$ coefficient. Upon squaring their ratio, we recover \eq{betovera}. We also recover here that the Hawking temperature $\kappa/2\pi$ is fixed by the late time exponential decay rate entering \eq{CarterL}. We finally notice that the stationarity of the flux is nontrivial. It follows from the fact that the ratio of \eq{betovera} is independent of $\omi$, and from the fact that $|\beta_{\omo, \omi}|^2 \propto 1/\omi$ for $\omi \to \infty$~\cite{Brout:1995rd}. 

\section{Emission from a universal horizon}
\label{sec:Horava}

\subsection{The model}

We aim to compute the late time radiation of a dispersive field propagating in a collapsing geometry. In principle, the radiation and the background fields should both obey the field equations of some extended theory of gravity, such as Ho\v{r}ava-Lifschitz gravity~\cite{Horava:2009uw} or Einstein-\ae ther theory~\cite{Jacobson:2000xp, Eling:2004dk}. Since our aim is to study the radiation rather than the collapse, the latter shall be described by a simplified model. At the end of the calculations, we shall argue that our results do not qualitatively rely on the particular model we use. 

For reasons of simplicity, we assume that the collapsing object is a null thin shell, and that the external geometry is still Schwarzschild. In this case, the metric is again given by \eq{eq:met}, and the Penrose diagram of \fig{fig:RelST} still covers the whole space-time. To describe the (unit time like) \ae ther field $u^\mu$ in the external region outside the shell, we adopt the solution of~\cite{Berglund:2012bu} (also used in~\cite{Cropp:2013sea}) with $c_{123}=0$, $r_0=2M$, and $r_u=0$.  The Killing horizon is still at $r=2M$, whereas the universal horizon, where $u^\mu K_\mu = 0$, is located at $r=M$. Inside the shell, we assume that the \ae ther field is at rest. To our knowledge, this configuration has not been shown to be a solution of the field equations. However, as explained in Subsection~\ref{sub:gen}, small deviations from this configuration should not significantly modify our conclusions.

In EF coordinates, on both sides of the shell, the \ae ther field $u^\mu$, and its orthogonal spacelike unit field $s^\mu$ are given by
\be \label{eq:uands}
&&u^\mu \pd_\mu = \pd_v - \frac{M(v)}{r} \pd_r, \nonumber\\
&&s^\mu \pd_\mu = \pd_v + \lp 1-\frac{M(v)}{r} \rp \pd_r,
\ee
where $M(v) = \Theta(v - 4M) M$. We introduce the ``preferred'' coordinates $t,X$ by imposing that $u_\mu dx^\mu \propto dt$ and $s^\mu \pd_\mu = {\rm sgn} (r-M) \pd_X$. Their precise definition is given in Appendix~\ref{app:PC}. In these coordinates, the metric takes the Painlev\'e-Gullstrand form: 
\be 
ds^2 = c^2 dt^2 - \lp dX - V dt \rp^2, 
\ee
where
\be \label{eq:vandc}
&&V = - \frac{M(v)}{r}, \nonumber \\
&&c = 
\left\lvert
K^\mu u_\mu 
\right\rvert = \left\lvert 1 - \frac{M(v)}{r} \right\rvert . 
\ee
At fixed $t$, outside the shell, $V$ and $c$ only depend on $X$. We notice that 
\be 
u_\mu dx^\mu = c dt. 
\ee
The factor $c$ ensures that $dt$ is a total differential. Moreover, as explained in Appendix~\ref{app:acc}, $c$ is constant when the \ae ther field is geodesic. Here we work with an accelerated \ae ther, which is a necessary condition to have a universal horizon. Importantly, $c$ vanishes on the universal horizon.~\footnote{In an analogue gravity perspective~\cite{Unruh:1980cg, Barcelo:2005fc}, to reproduce such a situation one needs a medium in which the group velocity of low-frequency waves vanishes at a point. From \eq{eq:DR}, we see that the effective dispersive scale $\Lambda / c$ must be divergent at the point where $c \to 0$. It would be interesting to find media which could approximatively reproduce this behavior.} In fact, the novelties of the present situation with respect to the standard case studied in~\cite{Macher:2009tw} only arise from the vanishing of $c$, and the associated divergence of the dispersive scale $\Lambda / c$.

\begin{figure}
\begin{center}
\includegraphics[scale=1]{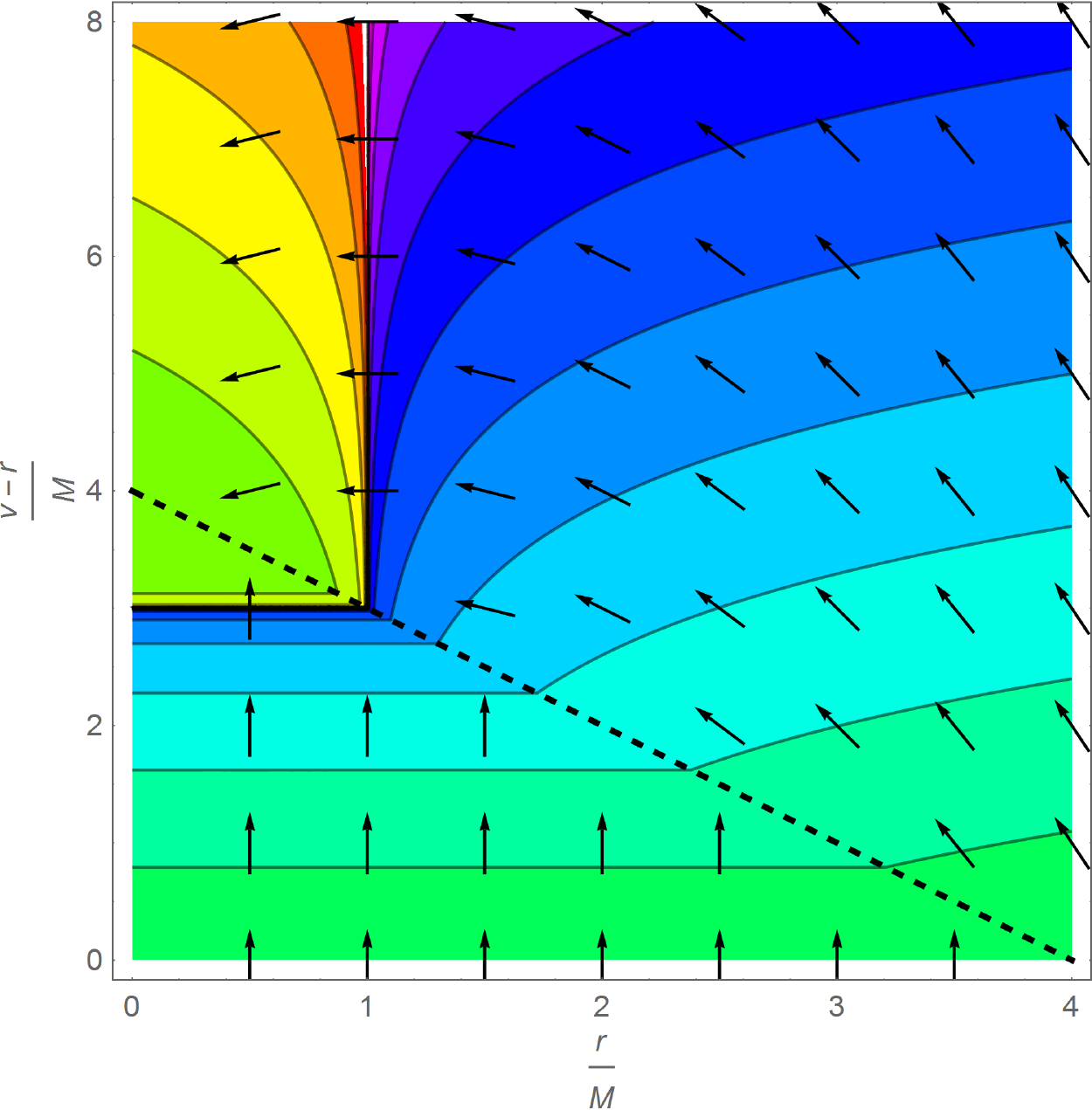} 
\caption{In this figure we show the lines of constant preferred time for the collapsing geometry in the plane $(v-r)/M,r/M$. The dashed line represents the trajectory of the null shell $v=4M$, and arrows show the direction of the aether field $u_\mu$. Notice that the external preferred time $t$ diverges on the universal horizon $r=M$, $v>4M$, whereas the internal time $T$, which is equal to $v - r$ inside the shell, covers the entire space-time. 
} \label{fig:pref}
\end{center}
\end{figure}

In \fig{fig:pref} we show the lines of constant preferred time, and the direction of the aether field $u_\mu$, in the $v,r$ plane. The coordinate $t$ is discontinuous across the shell trajectory, as was the null coordinate $u$ in the former section. As in the relativistic case, outside the shell we must use two coordinates $t$ and $t_L$, now on either side of the universal horizon. The inside coordinate $T$ evaluated along the shell, at $v=4M^-$, is a monotonically increasing function of both $t(v=4M^+,r)$ for $r > M$ and of $t_L(v=4M^+,r)$ for $r< M$. So, the foliation of the entire space-time by the inside coordinate $T$ is globally defined and monotonic. 

We consider a real massless dispersive field $\Phi$ with a superluminal dispersion relation. Its action is given by \eq{eq:relativisticaction} supplemented by a term quartic in derivatives:
\be \label{eq:action}
S = \int d^4x \sqrt{-g} \left[ \lp \pd_\mu \Phi \rp \lp \pd^\mu \Phi \rp - \frac{1}{\Lambda^2} \lp \nabla_\mu \lp h^{\mu \nu} \nabla_\nu \Phi \rp \rp \lp \nabla_\rho \lp h^{\rho \sigma} \nabla_\sigma \Phi \rp \rp\right], 
\ee
where $\nabla_\mu$ is the covariant derivative and $h^{\mu \nu} \equiv g^{\mu \nu} - u^\mu u^\nu$ is the projector on the hyperplane orthogonal to $u^\mu$. The dispersive momentum scale is given by $\Lambda$. The field equation reads
\be \label{fieldeq}
\nabla_\mu \nabla^\mu \Phi + \frac{1}{\Lambda^2} \lp \nabla_\mu  h^{\mu \nu}  \nabla_\nu \rp^2 \Phi = 0.
\ee
Using a $(1+1)$-dimensional approximation, \eq{fieldeq} reduces to 
\be \label{eq:2DFE}
\left[
\left[ \pd_t + \pd_\X V \right] \frac{1}{c} \left[ \pd_t + V \pd_\X \right] - \pd_\X c \pd_\X + \frac{1}{\Lambda^2} \pd_\X c \pd_\X \frac{1}{c} \pd_\X c \pd_\X
\right]
\psi = 0,
\ee 
when working in the preferred coordinates. Since this (self-adjoint) equation is second order in $\pd_t$, the Hamilton structure of the theory is fully preserved. In particular, the conserved scalar product has the standard form 
\be
\lp \psi_1 | \psi_2 \rp
= i \int dX \lp \psi_1^* \Pi_2 -  \Pi_1^* \psi_2 \rp, 
\label{scal}
\ee
where $\Pi = u^\mu \pd_\mu \psi = \lp \pd_t \psi + V \pd_X \psi \rp /c$ is the momentum conjugated to $\psi$. For more details; see Appendix~\ref{app:sc}. 

The Hamilton-Jacobi equation associated with \eq{eq:2DFE} is
\be \label{eq:DR}
{\Omega^2} = (\omo - V(X) P)^2 = {c(X)^2}\left[ P^2 + \frac{P^4}{\Lambda^2}\right]. 
\ee
We introduce the Killing frequency $\omo$, the preferred frequency $\Omega$, and the preferred momentum $P$: 
\be 
&& \omo = - K^\mu \partial_\mu S = - \pd_t S, \label{Kilf} \\ 
&& \Omega =  - 
c(X) \, u^\mu \pd_\mu S = \omo - V(X) P, \\
&& P  =  s^\mu \pd_\mu S = \partial_X S. 
\ee 
In these equations $S$ should be conceived as the action of a point particle; see~\cite{Brout:1995wp,Balbinot:2006ua,Coutant:2011in}. As explained in these works, $S$ governs the WKB approximation of the solutions of \eq{eq:2DFE}. Notice that \eq{Kilf} only applies outside the shell, whereas all the other equations make sense on both sides. 

\subsection{The modes and their characteristics}

To compute the late time radiation one should identify the various solutions of \eq{eq:2DFE}, and understand their behavior. In the presence of dispersion, one loses the neat separation of null geodesics into the outgoing $u$ ones, and the infalling $v$ ones. In what follows, we call $P^u$ ($P^v$) the roots of the dispersion relation which have a positive (negative) group velocity in the frame at rest with respect to the ``fluid'' of velocity $V$; see~\cite{Macher:2009tw}. Similarly, the corresponding modes will also carry the upper index $u$ or $v$. 

\subsubsection{The in and out asymptotic modes} 

In the internal region $v<4M$, the situation is particularly simple. Since the velocity field $V$ vanishes, the preferred frequency is $\omi = -\pd_T S$, and the dispersion relation \eq{eq:DR} becomes 
\be
\omi^2 = P^2 + \frac{P^4}{\Lambda^2}.
\ee
This relation is shown in the left panel of \fig{fig:DR}. At fixed $\omi$, the positive frequency modes with wave vectors $P^{u}(\omi) > 0$ and $P^{v}(\omi) < 0$ define the two \textit{in} modes $\phi^{u, \, \rm in}_\omi$ and $\phi^{v, \, \rm in}_\omi$. They both have a positive norm, which can easily be set to unity through a normalization factor. The mode $\phi^{u, \, \rm in}_\omi$ is the dispersive version of the relativistic in-mode of \eq{relat-in-mode}. 

Outside the shell, for $v>4M$, at fixed Killing frequency $\omo > 0$, the situation is more complicated as the number of real roots depends on $r$. Outside the Killing horizon, for $r>2M$, one has $c > \left\lvert V \right\rvert$. So, \eq{eq:DR} possesses two real roots $P^{u}(\omo) > 0$ and $P^v(\omo)< 0$, which describe outgoing and infalling particles, respectively. The WKB expression for the corresponding stationary modes [the solutions of \eq{eq:2DFE}] is 
\be \label{eq:WKBmodes}
\psi_\omo^{(i)} \lp t, X \rp \approx \frac{\exp \lp -i \lp \omo t  - \int^X P^{(i)}(\omo, X') dX' \rp \rp}{4 \pi\sqrt{\left\lvert \Omega(\omo,P^{(i)}) /( c(X) \, \pd_\omo P^{(i)}) \right\rvert}},
\ee
where $P^{(i)}(\omo, X)$ is a real solution of \eq{eq:DR} at a fixed $\omo$, and $\Omega(\omo,P^{(i)})$ the corresponding preferred frequency. These WKB modes generalize the expressions of \cite{Coutant:2009cu,Coutant:2011in} in that $c$ is no longer a constant. Using \eq{scal}, one easily verifies that they have a unit norm. One also verifies that the group velocity along the \textit{i}th characteristic is $dX^{(i)}/dt = 1/\pd_\omo P^{(i)}$. When considered far away from the black hole, $r/(2M) \gg 1$, the $u$-WKB mode is the dispersive version of the relativistic out-mode of \eq{relat-out-mode}. 

From this analysis, we see that there is no ambiguity to define the asymptotic behavior of the in and out modes, solutions of \eq{eq:2DFE}. As before, these two sets encode the black hole radiation through the overlaps of \eq{BogC}. To be able to compute these overlaps, we need to construct the globally defined modes. To this end, we must study both the behavior of $\psi_\omo^{(i)}$ near the horizon and the third kind of stationary modes which propagate in this region.

\subsubsection{Near horizon modes}

\begin{figure}
\includegraphics[width = 0.48 \linewidth]{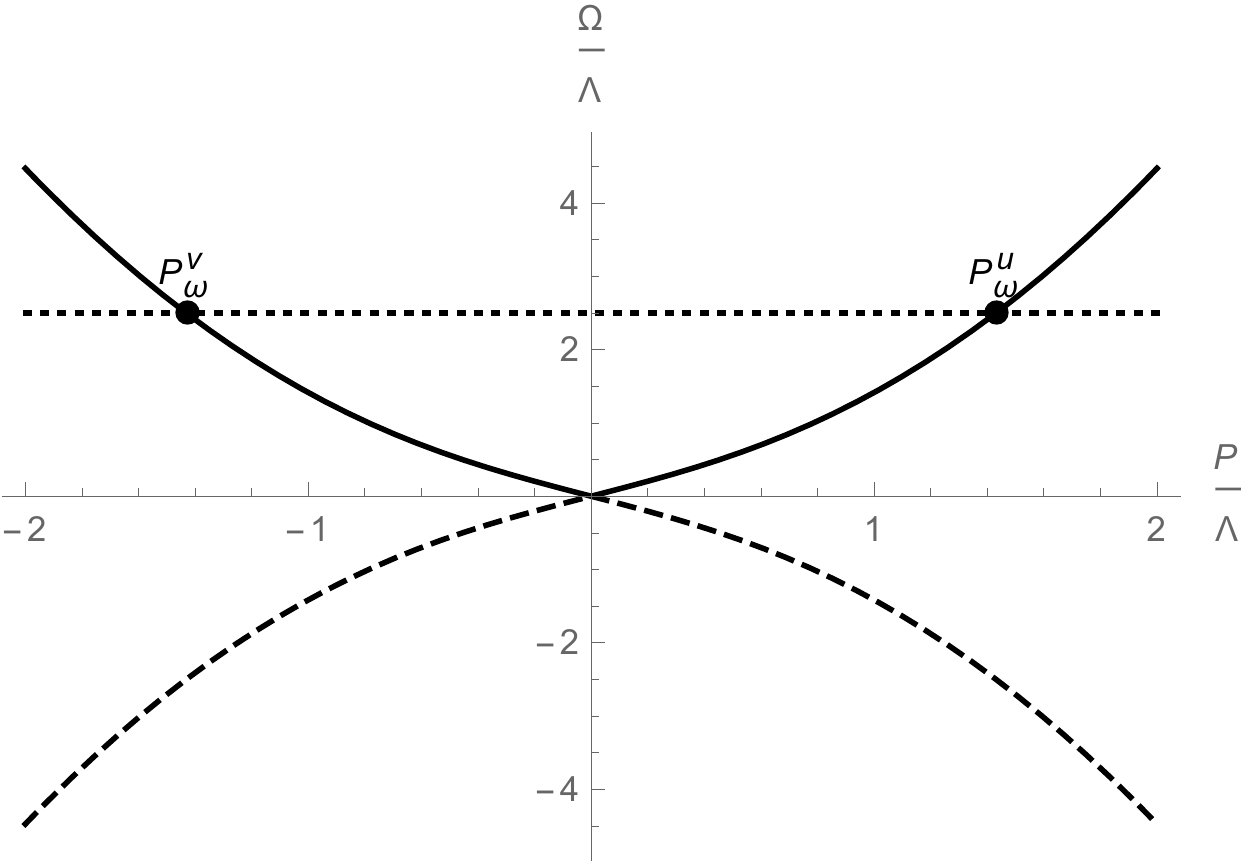}
\includegraphics[width = 0.48 \linewidth]{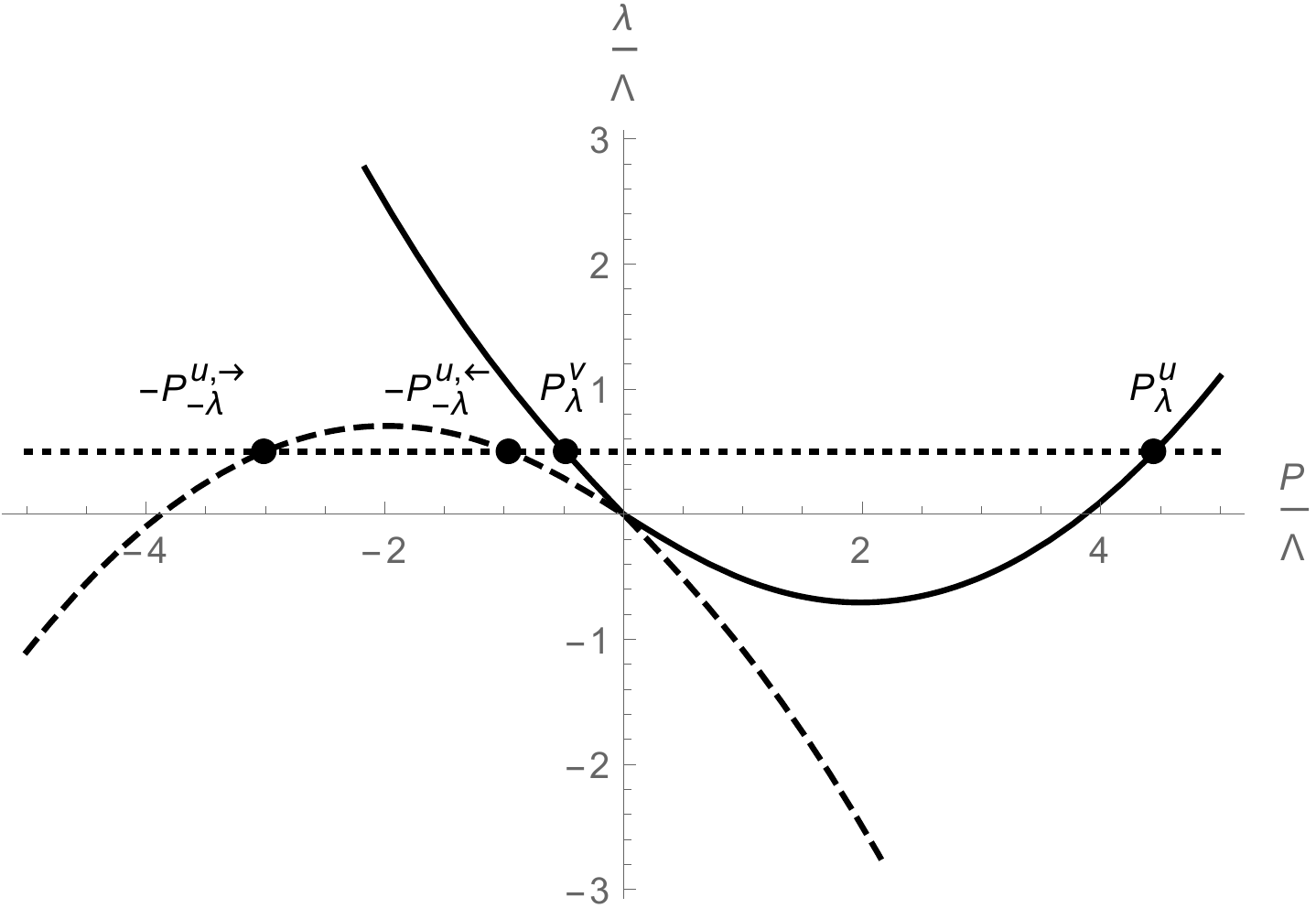} 
\caption{(Left panel) Dispersion relation in the internal region, where the preferred frame is at rest, in the $\Omega,P$ plane. The solid line shows $\omi = \Omega$ versus $P$ for the positive-norm modes. The dashed line corresponds to negative $\Omega$, i.e., negative-norm modes. The intersections with a line of fixed $\omi >0$ (dotted line) give the two solutions $P^{u}_\om$ and $P^{v}_\om$. Right: Dispersion relation in the ``superluminal'' region for $M < r < 2M$ in the $\omo , P$ plane. The two additional roots on the negative $\Omega$ $u$ branch are clearly visible. 
} \label{fig:DR}
\end{figure}

Inside the Killing horizon but outside the universal horizon, for $ M < r<2M$, one has $c < \left\lvert V \right\rvert$. As can be seen from the right panel of \fig{fig:DR}, one recovers the two roots $P^{u}(\omo) > 0$ and $P^{v}(\omo) < 0$ we just described. One notices that the $u$ root $P^{u}(\omo)$ has been significantly blueshifted, whereas the infalling root $P^{v}(\omo)$ hardly changed. Locally, in the WKB approximation, the corresponding modes are again given by \eq{eq:WKBmodes}. 

In addition, below a certain critical frequency $\lambda_c$ that depends on $c$ and $V$, we have two new real roots we call $-P^{(u, \rightarrow)}_{-\omo}$ and $-P^{(u, \leftarrow)}_{-\omo}$, where the arrow indicates the sign of the group velocity given by $1/\partial_\omo P$. (The minus signs in front of these roots and $\omo$ come from the fact that they have a negative preferred frequency $\Omega$ for $\omo > 0$. Hence, for $\omo = - \left\lvert \omo \right\rvert$, the mirror image roots, $P^{(u, \rightarrow)}_{-|\omo|}$ and $P^{(u, \leftarrow)}_{-|\omo|}$, have a positive $\Omega$.) Since $\Omega < 0$, the WKB modes associated with these roots have a negative norm~\cite{Coutant:2011in}. We call the right-moving one $\psir{-\omo}$ and the left-moving one $\psil{-\omo}$, so that the modes without complex conjugation have a positive norm. Both of them carry a negative Killing energy $- \omo$. Using Eqs. \eq{eq:WKBmodes} and \eq{scal}, one easily verifies that $\psir{-\omo}$ and $\psil{-\omo}$ have a negative unit norm within the WKB approximation. As we shall see they describe the negative energy partners trapped inside the Killing horizon before and after their turning point, respectively. 
To summarize the situation, it is appropriate to represent the characteristics of the three types of modes. We proceed as in~\cite{Brout:1995wp,Coutant:2011in}.

\subsubsection{The characteristics} 

\begin{figure}
\begin{center}
\includegraphics[width = 0.49\linewidth]{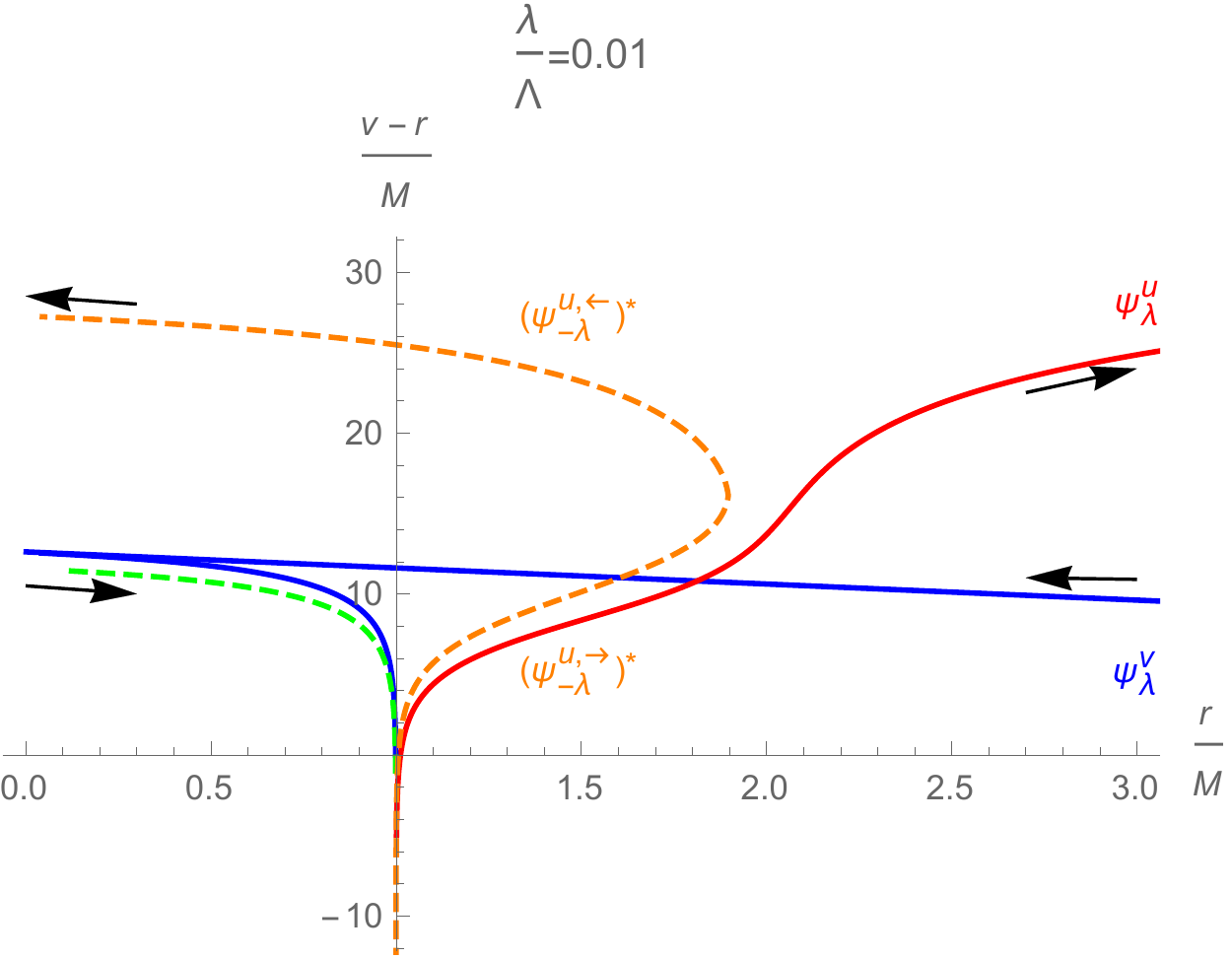} 
\includegraphics[width = 0.49 \linewidth]{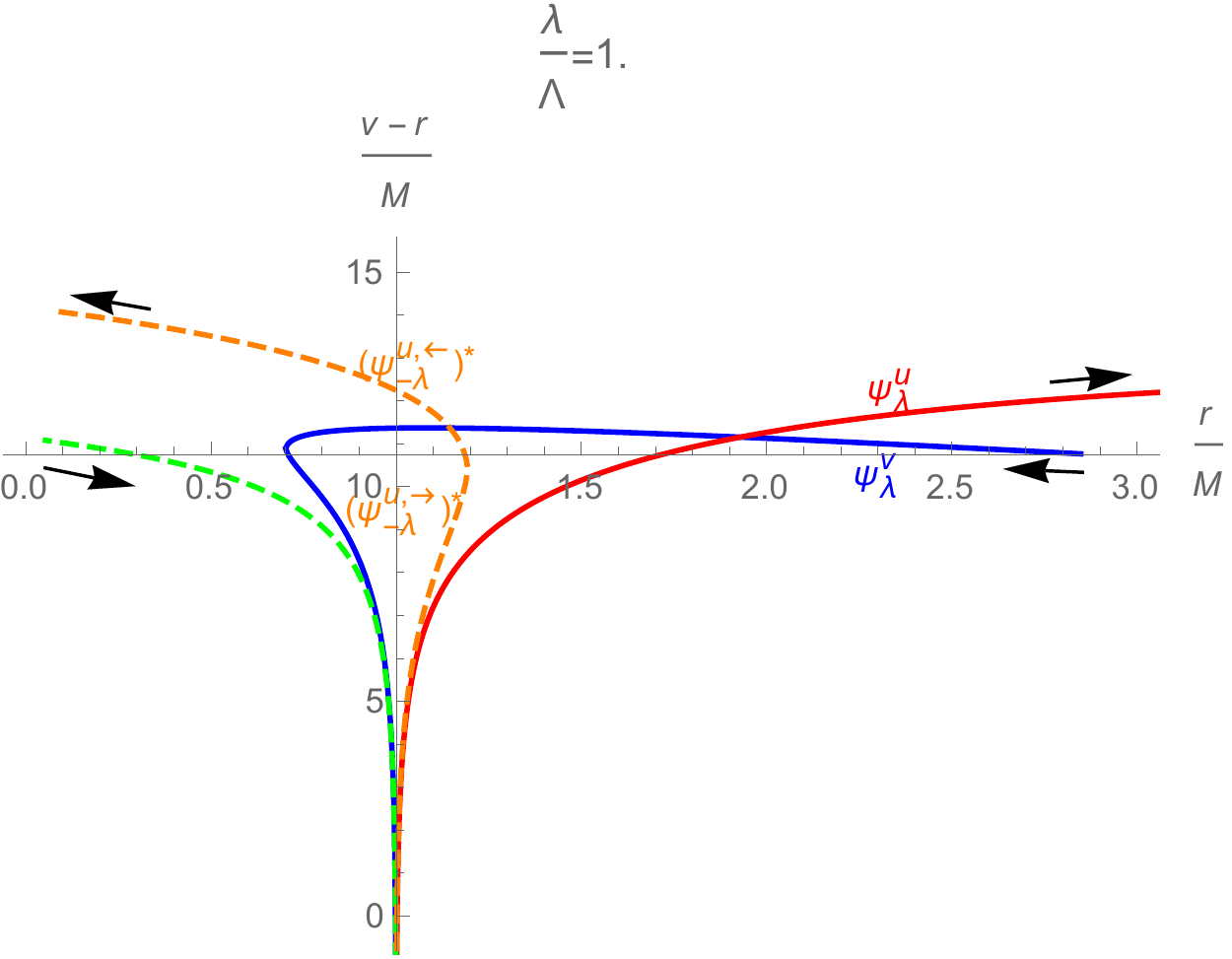}
\end{center}
\caption{Characteristics in a Schwarzschild stationary geometry for $\omo = 10^{-2} \Lambda$ (left) and $\omo = \Lambda$ (right). The arrows indicate the direction of increasing preferred time along each characteristic. Solid lines correspond to positive-norm modes, and dashed ones to negative-norm modes. For $r > M$, each characteristic is named by the corresponding mode. The green dashed line corresponds to an extra $v$ mode confined in $r <M$, as discussed in footnote~\ref{vfoot}. In this footnote, we also explain that the infalling $v$ mode (described by the blue line) possesses a turning point inside the universal horizon. The mode corresponding to the orange, dashed line is the high momentum WKB mode $\psir{-\omo}$ before the turning point, and the low momentum mode $\psil{-\omo}$  after it.  
} \label{fig:char}
\end{figure}

As said above, the characteristics are solutions of the Hamilton-Jacoby equation $\frac{dX}{dT} = \frac{1}{\partial_\omo P}$. Since the frequency is a constant of motion on each side of the mass shell, they can then be computed straightforwardly. In \fig{fig:char}, they are shown in the external region $v>4M$ for a small value of $|\omo |/ \Lambda = 0.01$ (left panel) and a moderate one $|\omo| / \Lambda = 1$ (right panel). The solid lines correspond to positive energy solutions while the dashed ones correspond to negative energy solutions.
 
The infalling $v$-like characteristics corresponding to $\psiv{\omo}$ (in blue) approach the universal horizon from infinity and cross it at a finite value of $v$. (When sending $\omo \to 0$ they asymptote to null infalling geodesics $v = cst.$) As their wave vectors are finite for $r \to M^+$, these characteristics will play no role in the sequel. As in the relativistic case, the $v$ modes act as spectators in the Hawking effect.~\footnote{Interestingly, $v$-like characteristics have a turning point inside the universal horizon $r<M$. (The presence of the turning point may be understood from the fact that, close to $r=0$, $|V|$ and $c$ go to infinity but $|V|/c$ goes to $1$. So, at fixed $\omo$ two roots merge at a point $r = r_{\rm tp}>0$. The turning point approaches $r=0$ in the limit $\omo \to 0$.) For later (preferred) times, they return towards the universal horizon, approach it asymptotically for $v \to -\infty$, $t_L \rightarrow + \infty$, and are highly blueshifted. In addition, for $r < M$, there is a new $v$ mode with negative norm for $\omo > 0$. It is indicated by a dashed green line in \fig{fig:char}. It emerges from the singularity and approaches the universal horizon while closely following the positive Killing frequency characteristic after its turning point. (In fact this new $v$ mode is directly related to the $u$ modes emerging from the singularity in~\cite{Jacobson:2001kz}: inside a universal horizon, $u$ and $v$ modes are swapped because of the vanishing of $c$ at $r=M$.) Since some of the $v$ modes originate from the singularity, and since the blueshift they experience is unbounded for $r \to M^-$, the $v$ part of the state will not obey Hadamard regularity conditions. This strongly indicates that the {\it inner} side of the universal horizon should be singular. This interesting question goes beyond the scope of the present paper.\label{vfoot}}

The $u$-like characteristics with positive energy (in red), corresponding to the WKB modes $\psiu{\omo}$, emerge from the universal horizon from its right ($r>M$) at early times. When $t$ increases, the momentum $P^u_\omo$ is redshifted while $r$ increases. At a finite time, the characteristics cross the Killing horizon, and go to infinity as $t \to \infty$ (almost along null outgoing geodesics when $\omo/\Lambda \ll 1$). 

The third characteristics (orange, dashed line) describe the trajectories followed by the negative-energy partners. For $t \to -\infty$, they also emerge from $r = M^+$. However, when increasing $t$ they have a turning point inside the Killing horizon, after which they move towards the universal horizon, smoothly cross it, and hit the singularity at $r=0$ at finite values of $v$ and $t_L$. Before the turning point, they are described by the WKB mode $\psir{-\omo}$, and after the turning point by $\psil{-\omo}$. 

It is important to notice that the {\it only} novel aspect with respect to the standard dispersive case (treated in full detail in~\cite{Coutant:2011in}) concerns the behavior near the universal horizon. To clarify these new aspects, we represent in \fig{fig:Gchar} the global structure of the characteristics in the collapsing mass shell geometry.

\subsubsection{The characteristics in the collapsing geometry}

In the internal region $v<4M$, the characteristics are straight lines. Coming backwards in time from the outside region, the inside trajectories are fixed by the value of the inside frequency which is determined (as in the relativistic case), by continuity of the field $\psi$ across the mass shell; see Appendix~\ref{app:match} for details. As a result, the derivative $\pd_r \psi$ must be continuous across $v = 4M$. At the level of the characteristics (i.e., in the geometrical optic approximation), this implies that $k_v$, the radial momentum at fixed $v$, is continuous along the shell. In terms of the inside and outside preferred momenta $P^u(\omi)$ and $P^u(\omo,r)$ evaluated at $v=4M^-$ and $v=4M^+$, respectively, the continuity condition gives
\be \label{eq:matching_char}
\left\lvert \frac{r}{r-M} \right\rvert \lp \omo + P^u(\omo,r) \rp = \omi+P^u(\omi).
\label{dCarterL}
\ee 
This equation has two solutions, but only one is well behaved as the other one gives a trajectory along which the preferred time is not monotonic. A straightforward calculation using the dispersion relation \eq{eq:DR} also shows that the sign of $\Omega$ is preserved. It should be noted that \eq{eq:matching_char} is the dispersive version of the relativistic equation $|r/(r - 2M)| \omo = \omi$, which gives back \eq{CarterL} for $r> 2M$, $\omi \gg \omo$, and when using $u$ rather than $r$. 

\begin{figure}
\begin{center}
\includegraphics[width=0.5\linewidth]{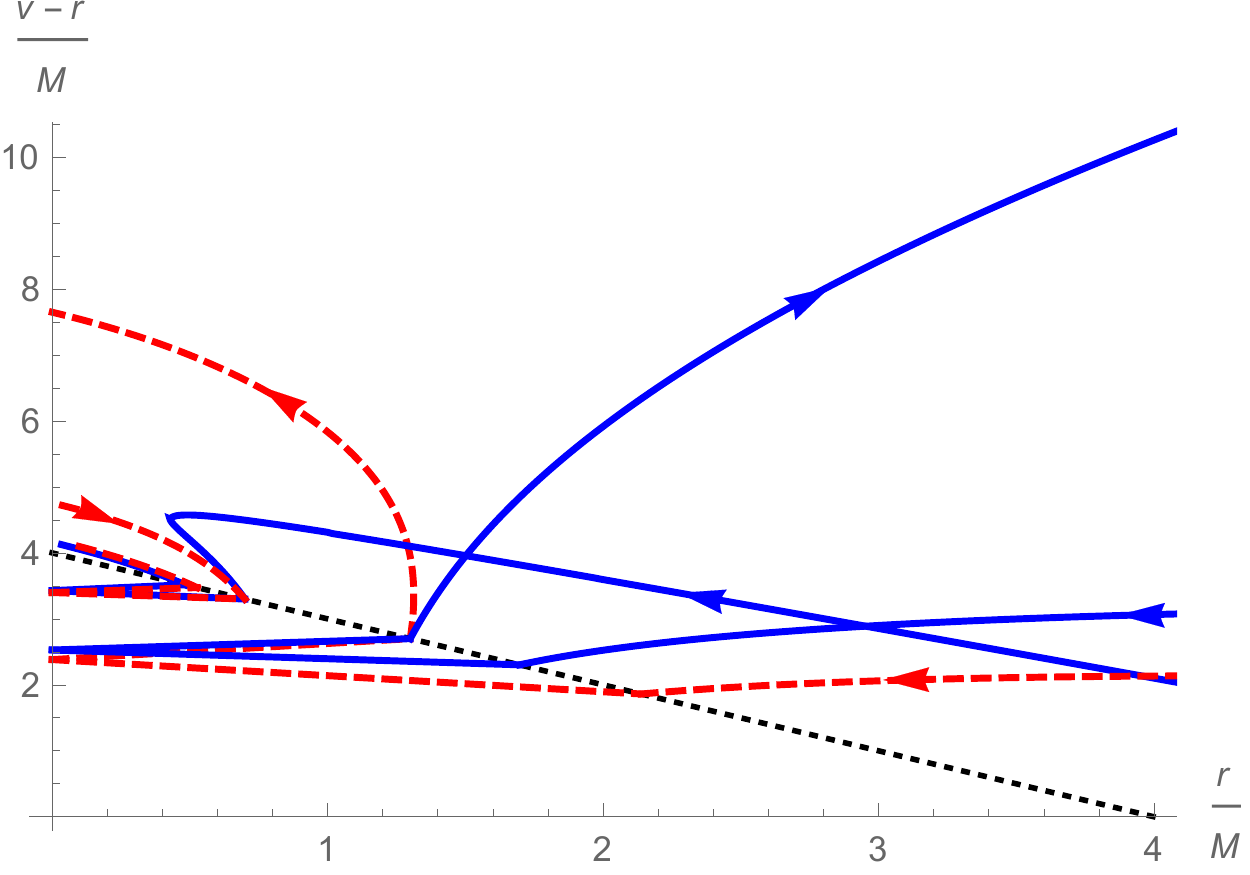} 
\end{center}
\caption{Characteristics crossing the infalling shell in the $v-r,r$ plane. The Killing frequencies of the outgoing $u$ modes and the incoming $v$ modes is $\omo = \pm 0.5 \Lambda$. The solid (dashed) lines represent characteristics for which the value of the Killing frequency $\omo$ of the out-going $u$ mode is positive (negative). The arrows indicate the future direction associated with the aether field. When tracing backwards the $u$-like characteristics associated with the Hawking quanta ($\omo > 0$) and their inside negative energy partners ($\omo < 0$),  we see that they both originate from infalling $v$-like superluminal characteristics with a high and positive Killing frequency $\lambda_{\rm in}$. The $v$ mode which emanates from the singularity (the dashed line) returns to it after having bounced at the center of the shell (not represented), closely following the characteristic of the $v$ mode coming from infinity which hits the shell at the same value of $r$. } \label{fig:Gchar}
\end{figure}

It should be also emphasized that all outgoing $u$-like characteristics originate from inside the shell, as in the relativistic case. This is shown in \fig{fig:Gchar}. Therefore, thanks to the universal horizon, the state of the field inside the shell determines the state of the $u$ modes. In this we avoid the problem discussed in~\cite{Jacobson:2001kz}, namely that in the {\it absence} of a universal horizon, the $u$ modes of a superluminal field originate from the singularity at $r = 0$. As discussed in  footnote~\ref{vfoot}, these modes still exist, but they are now trapped inside the universal horizon. 

Finally, we notice that the Killing frequency $\lambda_{\rm in}$ of the incoming $v$ modes which generate the outgoing $u$ modes exiting the shell at $r_c \approx M$ is very large. More precisely, when dealing with $u$ characteristics with positive $\Omega$ (i.e., modes with positive norm), irrespective of the sign of their Killing frequency $\omo$, the Killing frequency $\lambda_{\rm in}$ is positive. A straightforward calculation (based on the continuity of $k_v$ applied to the $v$ modes) shows that it scales as $\lambda_{\rm in} \approx 3 \Lambda M / \lp r_c - M \rp$.

For completeness, we have also represented in \fig{fig:Gchar} a couple of infalling $v$ characteristics which enter the shell for $ 0 < r < M$. One comes from $r = 0$ (the dashed line), and one from $r = \infty$ (the solid line). They both reach the singularity after having bounced at $r = 0$ inside the shell. These characteristics, although interesting, play no role in the Hawking process. 

\subsection{Behavior of the WKB modes near the universal horizon}

To be able to compute the late time behavior of the Bolgoliubov coefficients, we need to further understand the properties of the stationary modes in the immediate vicinity of the universal horizon at $r=M$. For $r>M$, the two roots $P_\omo^v$ and $P_{-\omo}^{u,\leftarrow}$ remain finite as $r \to M$. As can be seen in \fig{fig:char}, the associated trajectories smoothly cross the horizon. They thus play no role in the large $\omi$ limit. In fact they describe out modes. 

The two other roots $P_\omo^u$ and $P_{-\omo}^{u,\rightarrow}$ both diverge as $r\to M$. Importantly, they both satisfy 
\be 
 P^{\rm in}_{\pm \omo} = \frac{\Lambda M}{r - M} \pm \frac{r}{M} \omo  + \mathcal{O} \lp 1 - \frac{M}{r} \rp,
\label{Pomo}
\ee
where the + sign applies to $P_\omo^u$, and the - sign to $P_{-\omo}^{u,\rightarrow}$. We have added a superscript $in$ to emphasize that this behavior is relevant at early time $t$, just after having crossed the shell. The simple relation between $P_\omo^u$ and $P_{-\omo}^{u,\rightarrow}$ implies that for $r \to M$, the two WKB modes $\psi^u_{\omo}$ and $\psi_{-\omo}^{u, \rightarrow}$ are also related to each other by flipping the sign of $\lambda$. In the forthcoming discussion, to implement these points, we shall replace $\psi_{-\omo}^{u, \rightarrow}$ by $\psi^u_{-\omo}$, and add a superscript $in$ to the WKB modes $\psi^u_{\pm \omo}$. 

The appropriate character of this superscript can be understood as follows. Although the divergence in $1/ (r-M)$ in \eq{Pomo} resembles to what is found in the relativistic case, it has a very different nature due to the different relationship between $r$ and the preferred coordinate $X$. This can be seen by looking at the validity of the WKB approximation for $\psi^{u, \rm in}_{\omo}$ close to the universal horizon. Deviations from this approximation come from terms in $(\pd_X r) / (r P)$, $(\pd_X (r-M)) / ((r-M) P)$, and $(\pd_X P) / P^2$. 
Using 
\be
\pd_X = \frac{r-M}{r} \pd_r, 
\ee 
we find that these three terms go to zero as $r \to M$. Therefore, close to the universal horizon, the WKB approximation of~\eq{eq:WKBmodes} becomes exact for $\psi^{u, \rm in}_{\pm\omo}$. In fact, these modes behave as the dispersive $in$ modes near a Killing horizon~\cite{Brout:1995wp,Coutant:2011in}. Namely, they have a positive norm for all values of $\omo$ and, moreover, contain only positive values of $P^u$. We recall that this is the key property which also characterizes the so-called Unruh modes~\cite{Unruh:1976db, Brout:1995rd} for a relativistic field.

These are strong indications that no stationary emission should occur close to the universal horizon, as the pair production mechanism rests on deviations from the WKB approximation. This is confirmed in the next subsection. 

\subsection{Bogoliubov coefficients from the scattering on the shell}

We are now in a position to determine the scattering coefficients which govern the propagation across the null shell. Inside the shell, one has the $in$ mode $\phi^{u,\, {\rm in}}_{\omi}$. Along the shell, for $v = 4M^-$, it is a plane wave which behaves as $\phi^{u,\, {\rm int}}_{\omi} \sim \exp{[i (\omi + P^u(\omi)) r]}$. After having crossed the shell, for $r/M - 1 \ll 1$, it may be expanded in terms of the four WKB modes (which form a complete basis)
\be 
\phi^{u,\, {\rm in}}_{\omi} = \int^\infty_{-\infty} d \omo \lp \gamma_{\omi,\omo} \psi^{u, \rm in}_{\omo} + \delta_{\omi,\omo} (\psi^{u, \rm in}_{-\omo})^* + A_{\omi,\omo} \psiv{\omo} + B_{\omi,\omo} \psil{\omo} \rp.
\label{modem}
\ee
We are interested in the coefficients $\gamma_{\omi,\omo}$ and $\delta_{\omi,\omo}$ which multiply the two modes with divergent wave vectors and opposite norms. It should be pointed out that the integral over $\omo$ runs from $-\infty$ to $\infty$. The other two coefficients $A_{\omi,\omo}$ and $B_{\omi,\omo}$ multiply the two modes which remain regular across the universal horizon in the $(v,r)$ coordinates. They vanish in the limit $\omi \to \infty$.

The calculation of $\gamma_{\omi,\omo}$ and $\delta_{\omi,\omo}$ is straightforward in the $(v,r)$ coordinates; see Appendixes~\ref{app:alpha} and~\ref{app:beta}. For $|\omo|\lesssim \Lambda$, we find that their ratio decays as
\be \label{eq:bovera}
\left\lvert \frac{\delta_{\omi,\omo}}{\gamma_{\omi,\omo}} \right\rvert \mathop{=}_{\omi \to \infty} \mathcal{O} \lp  \frac{\sqrt{M \Lambda}}{\omi}\rp \times \exp \lp -2 M P^u(\omi)\rp ,
\ee
where $P^u(\omi) \sim \sqrt{\omi \Lambda}$ in the present high frequency regime. Equation \eq{eq:bovera} is the main result of the present work. Its meaning is clear: at late time, corresponding to the emission close to the universal horizon and thus to very large values of $P^u_\lambda \sim \Lambda/(r/M - 1)$; see \eq{Pomo}, the propagation across the shell induces no mode mixing between the inside in-mode $\phi^{u, \, {\rm in}}_{\omi}$ and the high momentum WKB mode with negative norm $(\psi^{u, \rm in}_{-\omo})^*$, irrespective of the value (and the sign) of $\omo$. As a result, outside the shell, the state of the field is stationary, and the vacuum with respect to the annihilation operators associated with $\psi^{u, \rm in}_{\omo}$ for $\omo \in (-\infty, \infty)$.~\footnote{This conclusion differs from that reported in~\cite{Berglund:2012fk}. We do not understand the procedure adopted there, which apparently implies that the leading term in \eq{Pomo} does {\it not} contribute to the ratio of \eq{eq:bovera}, thereby giving rise to a steady thermal radiation governed by the surface gravity of the universal horizon. Instead, the saddle point evaluation of $\delta_{\omi,\omo}$ performed in Appendix~\ref{app:beta} establishes that the leading term of \eq{Pomo} gives the exponential damping in $e^{- 2MP}$ of \eq{eq:bovera}.} It thus correspond to the $in$ vacuum as described in~\cite{Brout:1995wp,Coutant:2011in}.~\footnote{To be complete, one should propagate backwards in time the inside field configurations, and verify that they correspond to vacuum $v$-like configurations for $r \to \infty, t \to -\infty$. To verify this, we computed the scattering coefficients encoding a change of the norm of the $v$ modes when crossing the shell. We found that they also decrease exponentially in $\sqrt{\omi \Lambda}$ for $\omi \to \infty$. We also recall here that the Killing frequency of the $v$ modes engendering a stationary $u$ mode diverges as $\lambda_{\rm in} \approx 3 \Lambda M / \lp r_c - M \rp$, where $r_c$ is the radius when the $u$ mode exits the shell.}

\subsection{Genericness of \eq{eq:bovera}}
\label{sub:gen}

In this subsection, we distance ourselves from the model we considered to see how the above results may be affected. We first consider a modification of the mass shell trajectory close to the universal horizon. From the calculation of Appendixes~\ref{app:alpha} and~\ref{app:beta}, the factor $\exp \lp -2 M \sqrt{\Lambda \, \omi} \rp$ in \eq{eq:bovera} comes from the fact that the phase of the mode inside the mass shell is $\theta_{\rm int} \approx -  \omi T \approx \omi \lp r - v \rp$, while that of the mode outside the mass shell is $\theta_{\rm out} \approx \Lambda M / x$, where $x \equiv (r/M) -1$. At fixed $v$, we find that the stationary phase condition applied to $\theta_{\rm in} \pm \theta_{\rm out}$ (the upper sign applies to $\gamma$ while the lower sign applies to $\delta$) gives back the large frequency limit of \eq{dCarterL} with $x$ real for $\gamma$, while $x$ is purely imaginary for $\delta$, with a modulus $\sqrt{\Lambda / \omi}$. Let us now consider an arbitrary shell trajectory close to the universal horizon. We define an affine parameter $y$ along this trajectory. The possible saddle points are located where
\be 
\frac{d}{d y} \lp \omi T \mp \frac{\Lambda M }{x} \rp = 0,
\ee 
i.e.,
\be 
\omi \frac{dT}{dy} \pm \frac{\Lambda M}{x^2} \frac{dx}{dy} = 0.
\ee
So, the location of the saddle is
\be 
x^* = \sqrt{\mp M \frac{dx}{dT} \frac{\Lambda}{\omi}}. 
\ee
We get the same result as before, up to the factor $-M \frac{dx}{dT}$. Therefore, the ratio $\left\lvert {\delta_{\omi,\omo}}/{\gamma_{\omi,\omo}} \right\rvert$ is still suppressed by an exponential factor in $M \sqrt{\Lambda \, \omi}$, with a coefficient depending on the velocity of the mass shell when it crosses $r=M$. 

We now consider a generalization of the dispersion relation \eq{eq:DR} with higher-order terms. Specifically, we consider the dispersion relation
\be \label{eq:generalDR}
\frac{\Omega^2}{c^2} = \sum_{j=0}^N \frac{P^{2j}}{\Lambda_j^{2 (j-1)}}.
\ee  
Close to the universal horizon, the divergent wave vectors follow
\be \label{eqP}
P \approx \pm \Lambda_N \, x^{\frac{-1}{N-1}}.
\ee
As before, the coefficient $\gamma$ corresponds to $\pm = +$ in \eq{eqP}. The value of the saddle point is then real, and the exponential factor appearing in $\gamma_{\omi, \omo}$ has a unit modulus. Instead, for the coefficient $\delta$, corresponding to the minus sign in \eq{eqP}, the solutions of the saddle point equation are
\be 
x^* = \lp \frac{\omi}{\Lambda} \rp^{\frac{1-N}{N}} e^{i \pi \frac{1+2l}{N}}, \; l \in \mathbb{Z}.
\ee
Taking only the saddle points with negative imaginary parts, we find that $\delta_{\omi, \omo}$ is suppressed by a factor which is exponentially large in $\omi^{1/N}$. Interestingly, when using the inside spatial wave number $P^u(\om)$ rather than the inside frequency $\omi$, the norm of the coefficient $\delta_{\omi, \omo}$ always decreases as $\exp \lp -M A P^u(\om) \rp$ with $A > 0$, which means that it is the diverging character of $P^u(\omi)$ which guarantees that its sign does not flip when crossing the shell.

Similarly, the exponential factor suppressing $\delta_{\omi,\omo}$ is mildly affected by a change in the metric and/or the form of the \ae ther field, provided the inside wave vector remains smooth, whereas the outside one diverges as 
a power law for $r \to r_{U H}$, where $r_{U H}$ is the radius of the universal horizon. This should remain valid as long as there is no divergence (or cancellation) preventing us from defining preferred coordinates in which the dispersion relation takes the form of~\eq{eq:generalDR} close to the universal horizon. Indeed, the construction of Appendix~\ref{app:PC} can be easily extended to a generic space-time with a Killing vector $\chi$, endowed with a generic timelike, normalized \ae ther field $u^\mu$. 

\section{Conclusions}

We computed the late time properties of the Hawking radiation in a Lorentz violating model of a black hole with a universal horizon. To identify the appropriate boundary conditions for the stationary modes of our dispersive field, we worked in the geometry describing a regular collapse, and assumed that the inside state of the field is vacuum at (ultra) high inside frequencies $\omi \gg \Lambda$. We then computed the overlap along the thin shell of the outwards propagating inside positive norm modes, and the outside stationary modes. In the limit where the shell is close to the universal horizon, we show that the overlap between modes of opposite norms decreases exponentially in the radial momentum $P$. This result comes from the peculiar behavior of the momentum when approaching the universal horizon with a fixed Killing frequency; see \eq{Pomo}. Although this behavior was found in a specific model, we then argued that it will be found for generic (spherically symmetric) regular collapses and superluminal dispersion relations. 

As a result, irrespective of the model, at late time, the state of the outgoing field configurations is accurately described, for both positive and negative Killing frequencies, by the WKB modes with large positive momenta $P$ (and a positive norm). In this we recover the standard characterization of outgoing configurations in their vacuum state in the near horizon geometry. Indeed, the condition to contain only positive momenta $P$ prevails for both relativistic and dispersive fields in the vicinity of the Killing horizon. The present work, therefore, shows that this simple characterization still applies in the presence of a universal horizon. 

Once this is accepted, the calculation of the asymptotic flux is also standard, and shows that for large black holes the thermality and the stationarity of the Hawking radiation are, to a good approximation, both recovered. This suggests that the laws of black hole thermodynamics should also be robust against introducing high frequency dispersion.

As a corollary of the divergence of the radial momentum on both sides of the universal horizon, noticing that the inside configurations are blueshifted (towards the future), and that they have no common past with the outside configurations, it seems that the field state cannot satisfy any regularity condition across the universal horizon. It would be interesting to study the space of the field states, and determine whether some dispersive extension of the Hadamard condition can be imposed on the universal horizon. In the negative case, it seems that the universal horizon will be replaced by a spacelike singularity. 

\acknowledgements{We thank Xavier Busch for enlightening discussions about the causal structure of spacetimes with a universal horizon. We also thank Ted Jacobson, David Mattingly, and Sergei Sibiryakov for interesting comments and suggestions. This work received support from the French National Research Agency under the Program Investing in the Future Grant No. ANR-11-IDEX-0003-02 associated with the project QEAGE (Quantum Effects in Analogue Gravity Experiments).} 

\appendix

\section{Wave equation and Bogoliubov coefficients}
\label{app:details}

In this appendix, we give the general formulas and main steps in the derivation of the results presented in Section~\ref{sec:Horava}.

\subsection{Preferred coordinates}
\label{app:PC}

The preferred coordinates $(t,\X)$ are defined by the followng four conditions
\begin{itemize}
\item $s^\mu \pd_\mu = \pm \pd_\X$ at fixed $t$;
\item $\pd_v = \pm \pd_t$ at fixed $X$;
\item $\pd_r T < 0$ along the shell trajectory;
\item $\pd_r \X > 0$ along the shell trajectory.
\end{itemize}
These four conditions uniquely define $t$ and $\X$ as
\be 
t = \left\lbrace
\begin{array}{cc}
v-r, & v<4M ,\\
v- r^*_U, 
 & v>4M \wedge r>M ,\\
- \lp v- r^*_U \rp,
& v>4M \wedge r<M,
\end{array}
\right.
\ee
and
\be 
\X = \left\lbrace
\begin{array}{cc}
r, & v<4M, \\
r^*_U, 
& v>4M \wedge r>M ,\\
- r^*_U, 
& v>4M \wedge r<M .
\end{array}
\right. 
\ee
In these expressions, $r^*_U = r + M \ln | \frac{r}{M}-1 |$ is the tortoise coordinate built around the universal horizon. 

\subsection{Wave equation and scalar product}
\label{app:sc}

The action \eq{eq:action} has a $U(1)$ invariance under $\Phi \to e^{i \theta} \Phi$, from which we derive the conserved current density
\be \label{eq:current}
\mathcal{J}^\mu \equiv -i \sqrt{-g} \lp \Phi \nabla^\mu \Phi^* - \frac{1}{\Lambda^2} h^{\mu \nu} \lp \nabla_\nu \Phi \rp \lp \nabla_\rho h^{\rho \sigma} \nabla_\sigma \Phi^* \rp + \frac{1}{\Lambda^2} \Phi h^{\mu \nu} \nabla_\nu \nabla_\rho h^{\rho \sigma} \nabla_\sigma \Phi^* \rp + c.c.,
\ee
where ``$c.c.$'' stands for the complex conjugate, satisfying
\be 
\pd_\mu \mathcal{J}^\mu = 0.
\ee
As the wave equation \eq{fieldeq} is linear, one easily shows that $\mathcal{J}^\mu$ defines a conserved (indefinite) inner product in the following way. Considering two solutions $\Phi_1$ and $\Phi_2$ of \eq{fieldeq}, we first define $\mathcal{J}^\mu \lp \Phi_1, \Phi_2 \rp$ by replacing $\Phi^*$ by $\Phi_1^*$ and $\Phi$ by $\Phi_2$ in \eq{eq:current}. The inner product of these two solutions is then defined by
\be \label{eq:inner}
\lp \Phi_1, \Phi_2 \rp_\tau \equiv \int d^3x \, n_\mu \mathcal{J}^\mu \lp \Phi_1, \Phi_2 \rp,
\ee
where $n_\mu$ is the unit vector perpendicular to the 3-surface defined by $\tau = cst$,
and $\tau$ is a time coordinate. When considering the 3-surfaces defined by $t= cst.$, the above overlap simplifies and gives the standard (Hamiltonian) conserved scalar product of \eq{scal}. 

\subsection{Matching conditions on the mass shell}
\label{app:match}

In order to compute the overlap of two modes defined on either side of the mass shell, we need the matching conditions to propagate the modes from the internal region to the external one and vice versa. As we now show, they appear naturally when considering the behavior of $\mathcal{J}^v \equiv \mathcal{J}^\mu \partial_\mu v$ across the shell. To see this, we first rewrite $\mathcal{J}^v \lp \Phi_1, \Phi_2 \rp$ as
\be \label{eq:innerbis} 
\mathcal{J}^v \lp \Phi_1, \Phi_2 \rp = && -i \lp \Phi_2 \sqrt{-g} \lp \nabla^v + \frac{1}{\Lambda^2} h^{v \mu} \nabla_\mu \nabla_\rho h^{\rho \sigma} \nabla_\sigma \rp \Phi_1^* 
- \frac{1}{\Lambda^2} \sqrt{-g} \lp h^{v \mu} 
\nabla_\mu \Phi_2 \rp \lp \nabla_\rho h^{\rho \sigma} \nabla_\sigma \Phi_1^* \rp \rp
\nn
&&- \lp \Phi_1^* \leftrightarrow \Phi_2 \rp. 
\ee
Inspecting \eq{fieldeq} and requiring that the second term has no singularity which cannot be canceled by the first one, we find that the quantities
\begin{itemize}
\item $\Phi$,
\item $\sqrt{-g} h^{0 \nu} \nabla_\nu \Phi$,
\item $\nabla_\rho h^{\rho \sigma} \nabla_\sigma \Phi$, and
\item $\sqrt{-g} \lp \nabla^0 + \frac{1}{\Lambda^2} h^{0 \nu} \nabla_\nu \nabla_\rho h^{\rho \sigma} \nabla_\sigma \rp \Phi$
\end{itemize}
are continuous across $v=4M$. Since the complex conjugate of a solution of \eq{fieldeq} is still a solution, this applies to $\Phi = \Phi_1^*$ as well as $\Phi = \Phi_2$. Therefore, in evaluating \eq{eq:innerbis} one can evaluate $\Phi_1^*$ and the operators acting on it on one side of the shell,  $v = 4M - \epsilon$, $\epsilon \to 0$, while $\Phi_2$ and the operators acting on it are evaluated on the other side $v = 4M + \epsilon$.

\subsection{Calculation of \texorpdfstring{$\gamma_{\omi,\omo}$}{}} 
\label{app:alpha}

Let us consider two radial modes known on different sides of the mass shell: $\Phi_1$ is known for $v<4M$ and $\Phi_2$ for $v>4M$. The complete expression of the scalar product in the $v,r$ coordinates is somewhat cumbersome, but it greatly 
simplifies in the relevant limit where
\begin{itemize}
\item $\psi_1$ has a large frequency $\left\lvert \omi \right\rvert \gg \Lambda$;
\item $\psi_2$ has a large wave vector $\left\lvert k_{v,2} \right\rvert \gg \omo, \Lambda$.
\end{itemize}
We have introduced the wave vector $k_v \equiv \pd_r S$ at a fixed $v$. For the modes we are interested in $k_{v,2} = \lp \omo_2 + P_2 \rp / x$ and $\omi$ are of the order $\Lambda/x^2$, where $x = (r/M)-1$. Keeping only the leading terms in the inner product then gives
\be 
\lp \Phi_1, \Phi_2 \rp_v \approx \frac{4 i \pi}{\Lambda^2} \int dr \lp 
-\psi_2 (\pd_v+\pd_r)^3 \psi_1^* + \lp 1- \frac{M}{r} \rp \pd_r \psi_2 \lp \pd_v+\pd_r \rp^2 \psi_1^* 
\right. \nn \left.
+ \psi_1^* \lp 1- \frac{M}{r} \rp^3 \pd_r^3 \psi_2 - \lp \pd_v + \pd_r \rp \psi_1^* \lp \frac{M}{r} -1 \rp^2 \pd_r^2 \psi_2,
\rp
\ee
with relative corrections of order $x$. When choosing for $\psi_1$ the $in$ mode of frequency $\omi$, and for $\psi_2$ the stationary WKB mode of \eq{eq:WKBmodes} with the large momentum given by \eq{Pomo}, we get
\be 
\lp \Phi_1, \Phi_2 \rp_v &\approx & 4 \pi M \int_{x>0} dx \lp \frac{P_{\omi}^3}{\Lambda^2} \pm \frac{P_{\omi}^2}{\Lambda x} \pm \frac{\Lambda}{x^3} + \frac{P_\om}{x^2} \rp \psi_1^* \psi_2 \nn
& \approx &  \frac{M e^{4 i M \lp \omi - \omo \rp} e^{-i M (\omi + P_{\omi})}}{4 \pi \sqrt{\Lambda \left\lvert \omi \lp \frac{d \om}{dP} \rp_1 \right\rvert}} \int_{x>0} dx \lp \frac{P_{\omi}^3}{\Lambda^2} \pm \frac{P_{\omi}^2}{\Lambda x} \pm \frac{\Lambda}{x^3} + \frac{P_\om}{x^2} \rp 
\nn
& & \exp 
\lp i \lp \mp \frac{\Lambda M}{x} + \lp 2 \omo \pm \Lambda \rp M \ln \left\lvert x \right\rvert - M \lp \omi + P_{\omi} \rp x \rp
\rp .
\ee
In this equation, as well as in the remainder of this appendix, the sign $\pm$ discriminates between $\gamma$ and $\delta$; see below. In the large frequency limit, we evaluate this integral through a saddle point approximation. The possible saddle points are the values of $x$ where 
\be 
\frac{d}{dx} \lp \mp \frac{\Lambda M}{x} -M \lp \omi + 
P_{\omi} \rp x \rp \approx \frac{d}{dx} \lp \mp \frac{\Lambda M}{x} -M \omi x \rp = 0,
\ee
i.e.,
\be 
x^2 \approx \pm \frac{\Lambda}{\omi}.
\ee
This is very similar to the saddle point condition applied to the Bogoliubov coefficients describing the scattering of plane waves on a uniformly accelerated mirror~\cite{Obadia:2002ch, Obadia:2002qe}.

The coefficient $\gamma_{\omi, \omo}$ is defined for $\pm \omi > 0$. Since the integral runs over $x>0$, we must choose the saddle point $x^*$ at
\be 
x^*_\gamma \approx \sqrt{\frac{\Lambda}{\left\lvert \omi \right\rvert}}.
\ee
We get
\be 
\gamma_{\omi, \omo} \approx \pm \sqrt{\frac{\mp i M}{2 \pi \left\lvert \omi \right\rvert}} \exp \lp i M \lp 3 \omi - 4 \omo -P_{\omi} \mp 2 \sqrt{\Lambda \left\lvert \omi \right\rvert} +\frac{1}{2} \lp 2 \omo \pm \Lambda \rp \ln \lp \frac{\Lambda}{\left\lvert \omi \right\rvert} \rp \mp \Lambda \rp \rp .
\ee
It is easily shown that, under these approximations, the following unitarity relation is satisfied:
\be 
\int_0^\infty d\omi \gamma_{\omi,\omo}^* \gamma_{\omi,\omo'} \approx \delta (\omo - \omo').
\ee
This implies that the $\delta_{\omi,\omo}$ coefficients are suppressed in the limit $\omi \to \infty$.

\subsection{Calculation of \texorpdfstring{$\delta_{\omi,\omo}$}{}} 
\label{app:beta}

The calculation of $\delta_{\omi,\omo}$ follows the same steps. The saddle point equation now is
\be 
x^{*2}_\delta  = -\frac{\Lambda}{\left\lvert \omi \right\rvert}. 
\ee
To be able to deform the integration contour to include the saddle point, we must choose the solution in the half-plane where the exponential decreases, i.e.,
\be 
x^*_\delta = - {\rm sgn} \lp \omi \rp i \sqrt{\frac{\Lambda}{\left\lvert \omi \right\rvert}}.
\ee
The exponential factor in the integral then gives a suppression factor 
\be 
\exp \lp -M \lp 2 \sqrt{\Lambda \left\lvert \omi \right\rvert} + \pi \lp -{\rm sgn} \lp \omi \rp \omo + \frac{\Lambda}{2} \rp + \Lambda \rp \rp.
\ee

In addition, to the order to which the calculation was performed, the prefactor vanishes. As the first relative corrections from neglected terms are of order $\mathcal{O}(x^*) = \mathcal{O}(\sqrt{\Lambda/\left\lvert \omi \right\rvert})$, we get 
\be 
\delta_{\omi,\omo} = \mathcal{O} \lp \frac{\sqrt{M \Lambda}}{\left\lvert \omi \right\rvert} \rp \exp \lp -2 M \sqrt{\Lambda \left\lvert \omi \right\rvert} \rp.
\ee 

\section{Acceleration of the \ae ther field}
\label{app:acc}

The acceleration of the \ae ther field is
\be 
\gamma^\mu = u^\nu \nabla_\nu u^\mu.
\ee
Using \eq{eq:uands}, this gives for $v \neq 4M$ 
\be 
\gamma^\mu \gamma_\mu = -\frac{M(v)^2}{r^4}.
\ee

For completeness, we now show that in $1+1$ dimensions a stationary universal horizon requires that the \ae ther field has a nonvanishing acceleration, thereby generalizing what was found in de Sitter in~\cite{Busch:2012ne}. We consider a stationary space-time with Killing vector $K^\mu$, endowed with a timelike \ae ther field $u^\mu$. The universal horizon is defined as the locus where $K^\mu u_\mu = 0$. (Notice that the Killing field must thus be spacelike on the universal horizon.) In particular, $K^\mu$ cannot be aligned with $u^\mu$. Using the Killing equation, the variation of $K^\mu u_\mu$ along the flow of $u^\mu$ is
\be 
u^\mu \nabla_\mu \lp K^\nu u_\nu \rp = K_\mu \gamma^\mu.
\ee
If $u^\mu$ is freely falling, $\gamma^\mu = 0$ and $u^\mu$ is tangent to the hypersurfaces of constant $K^\mu u_\mu$. In particular, it is tangent to the universal horizon. In $1+1$ dimensions, since $K^\mu$ and $u^\mu$ cannot be aligned, $K^\mu$ is not a tangent vector to the universal horizon, which is thus not stationary. Models with a stationary universal horizon are thus in a different class than those studied in~\cite{Coutant:2011in}.

\begin{figure}
\includegraphics[width=0.48 \linewidth]{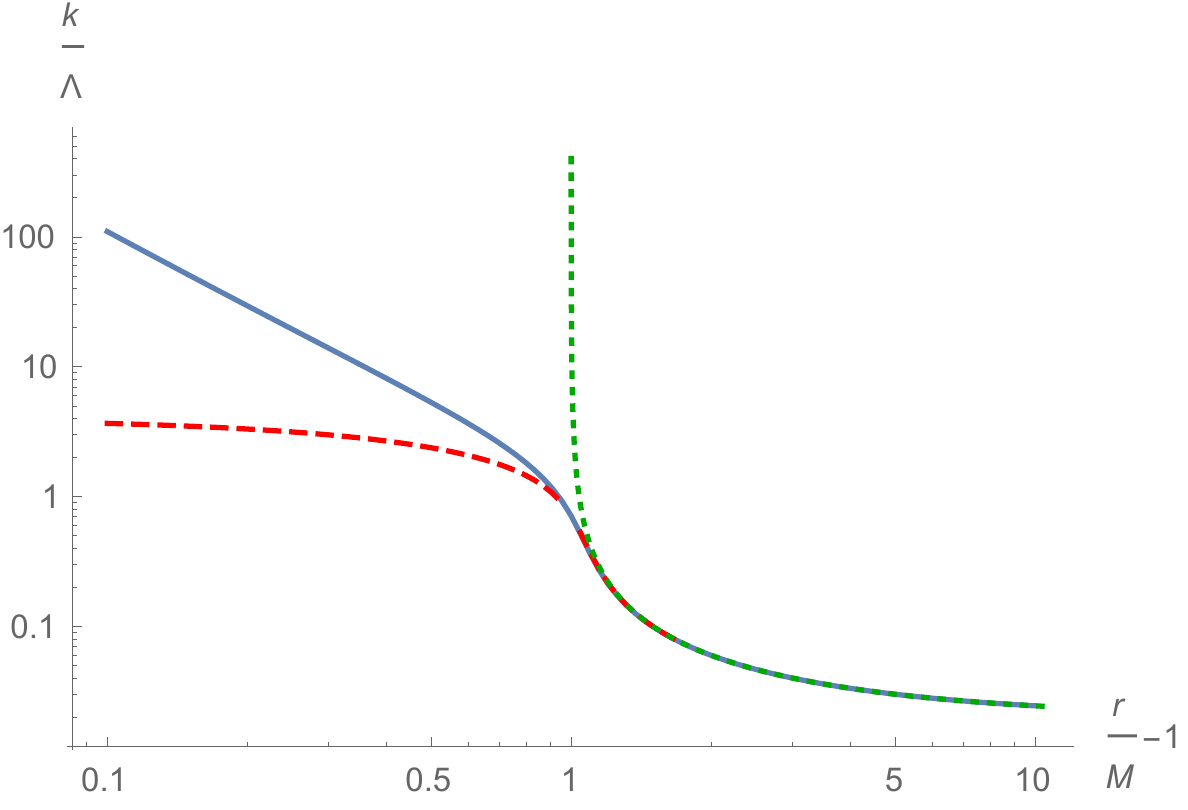}
\caption{Comparison of three wave vectors $k(r,\omo)/\Lambda$ as a function of $x = r/M - 1$ (both in logarithmic scales) in the relativistic case (green, dotted line), for freely falling preferred frame (red, dashed line), and in the model of Section~\ref{sec:Horava} (blue, solid line). The Killing frequency is $\omo = 10^{-2} \Lambda$, and the Killing horizon is located at $x = 1$. One clearly sees the unbounded growth of the relativistic wave number. More importantly, one also sees that the two dispersive wave vectors behave in the same manner across the Killing horizon. Hence the acceleration of $u^\mu$ has a significant effect on $k$ only when approaching the universal horizon. } \label{fig:HvsC}
\end{figure}

To see the combined effects of the dispersion and acceleration, we show in \fig{fig:HvsC} the local value of the wave vector in the $v,r$ coordinates, $k_v$, for the outgoing $u$ mode, as a function of $r$. We compare three models with the same parameters, and for $\omo = 10^{-2} \Lambda$. The blue, solid curve shows the result for the model of Section~\ref{sec:Horava}. The green, dotted curve shows the relativistic case. The red, dashed one shows the result for a dispersive model with a nonaccelerated preferred frame chosen to coincide with the \ae ther frame of Section~\ref{sec:Horava} at $r=2M$.\footnote{It must be noted that this model is not well defined for $r \to \infty$. The reason is that at $r=2M$, we have $u \cdot \pd_v = 1/2$. A nonaccelerated vector field $w$ which coincides with $u$ at $r=2M$ must thus satisfy the two conditions $w \cdot \pd_t = 1/2$ and $w \cdot w = 1$ at $r=2M$. From the free-fall condition, these two properties extend in the whole domain where the preferred frame is defined. Since they are incompatible in Minkowski space, we deduce that the domain in which the preferred frame can be defined does not extend to $r \to \infty$. A straightforward calculation shows that it extends up to $r = 8 M /3$. However, as this model is well defined close to and inside the Killing horizon, it can be used to see the qualitative differences between the nonaccelerated and accelerated cases.} We see in \fig{fig:HvsC} that the three models give very similar results for $r<2M$. Close to $r=2M$, the relativistic wave vector diverges, while the nonaccelerated dispersive model still closely follows the accelerated one. When $r$ is further decreased, the predictions of the two models separate: the nonaccelerated one gives a finite wave vector at $r=M$ while the accelerated one gives $k \propto \lp r-M \rp^{-2}$. 

\section{Hawking radiation in the presence of a universal horizon}
\label{app:Hawking}

In Section~\ref{sec:Horava}, we showed that the late time emission from the universal horizon is governed by Bogoliubov coefficients which are exponentially suppressed when the inside frequency $\omi \gg \omo$. This result was obtained using the WKB approximation of the stationary modes just outside the universal horizon. This approximation is trustworthy as we verified that the deviations from the WKB treatment go to zero when approaching the universal horizon. This implies that at late time the inside vacuum is adiabatically transferred across the shell. Hence, the $u$ part of the field state can be accurately described by the WKB high (preferred) momentum mode $ \psi^{u, \rm in}_{\omo}$ for both signs of $\omo$. In this we recover the situation described in~\cite{Brout:1995wp,Balbinot:2006ua,Coutant:2011in}. Therefore, the nonadiabaticity that will be responsible for the asymptotic radiation will be found in the propagation from the universal horizon to spatial infinity. The value of the Bogoliubov coefficients should essentially come from the stationary scattering near the Killing horizon. Hence, we expect to get a nearly thermal spectrum governed by the surface gravity of the  Killing horizon, and with deviations in agreement with those numerically computed in~\cite{Finazzi:2012iu,Robertson:2012ku}.

To verify this conjecture, we numerically propagate the outgoing mode $\phiu{\omo}$ from a large value of $r/2M$ down inside the trapped region to $r \to M^+$. This mode can be written in the limits $r \to M$ and $r \to \infty$ as 
\be
&& \phiu{\omo}(r) \mathop{\sim}_{r \to \infty} \psiu{\omo} + A_\omo \psiv{\omo}, \nn
&& \phiu{\omo}(r) \mathop{\sim}_{r \to M} \alpha_\omo \psi^{u, \rm in}_{\omo} + \beta_\omo (\psi^{u, \rm in}_{-\omo})^*, 
\ee 
where the WKB modes are as described in Section~\ref{sec:Horava}. The coefficient $A_\omo$ governs the grey body factor. In our (1+1)-dimensional model, we have verified that it plays no significant role. (We found that $|A_\omo|^2$ is bounded by $0.16$.) Hence, as usual, the Hawking effect is essentially encoded in the mode mixing of $u$ modes of opposite norms.  

\begin{figure}
\includegraphics[width=0.48 \linewidth]{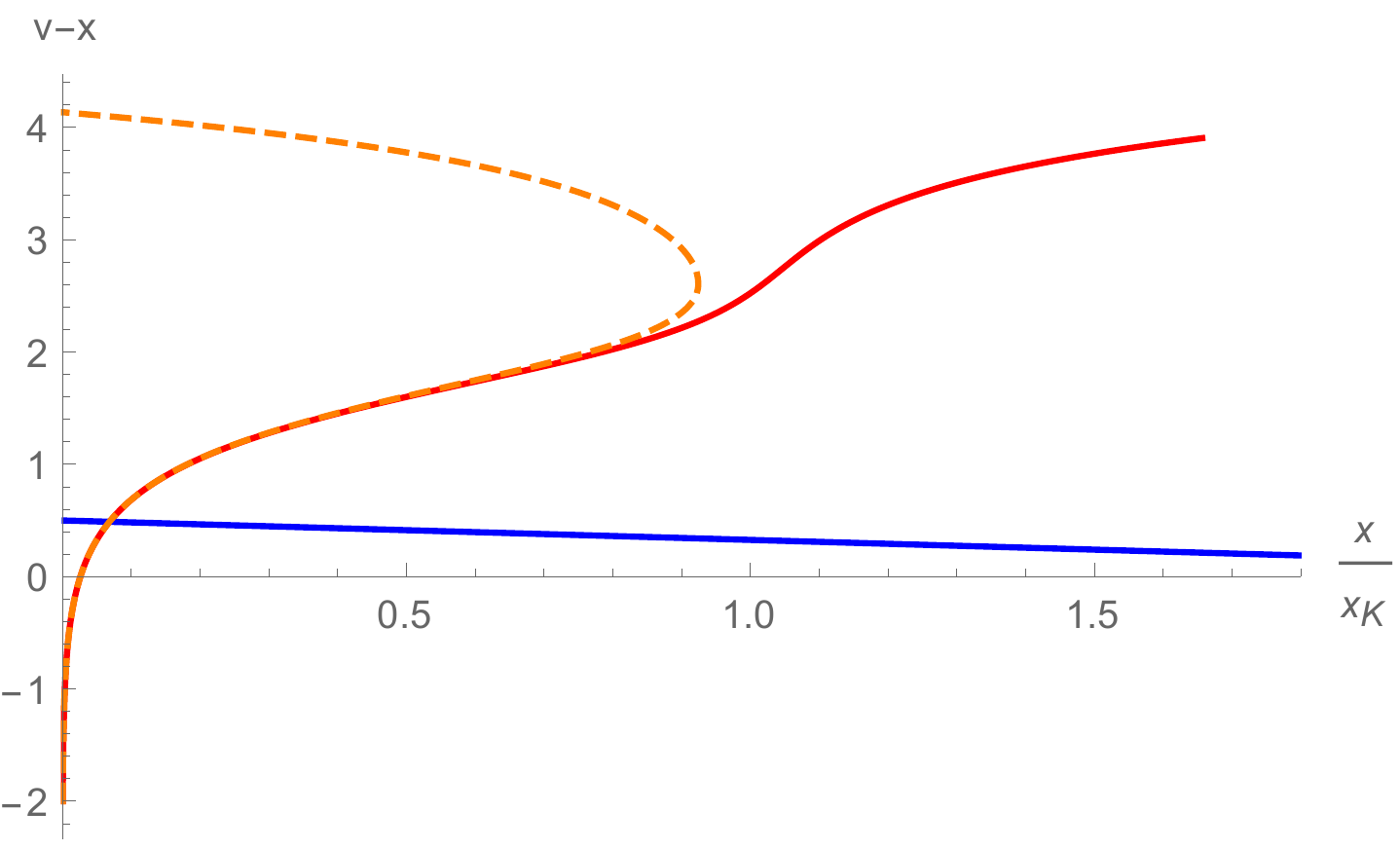} 
\includegraphics[width=0.48 \linewidth]{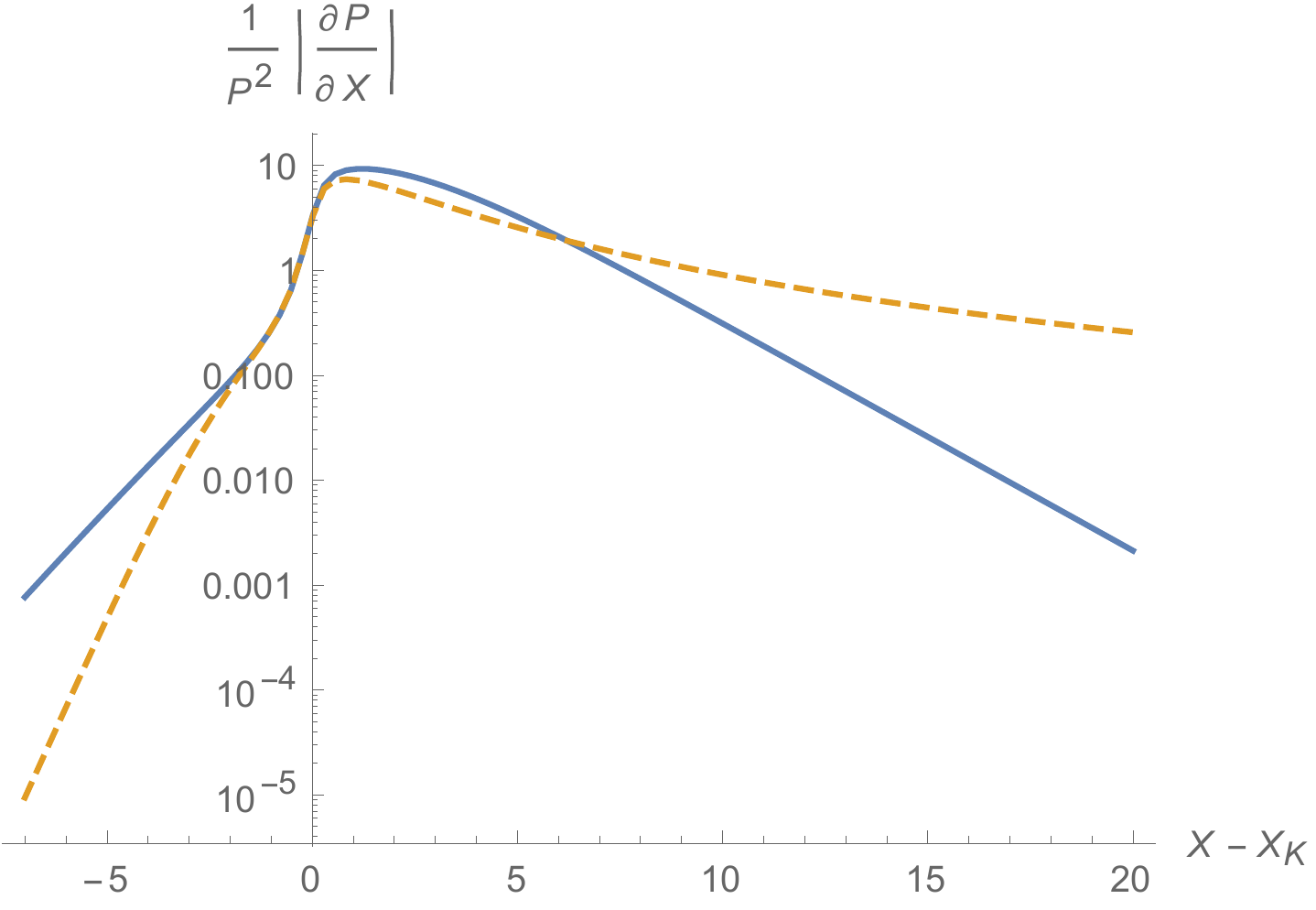}
\caption{(Left panel) Characteristics in the $v-x,x/x_K$ coordinates, where $x \equiv r-r_U$ and $x_K$ gives the location of the Killing horizon. Only the region outside the mass shell and the universal horizon is represented. The parameters are $\Lambda/\kappa \sim 1.1$, $X_0 = 0.5$, and $\lambda/\Lambda =0.01$. (Right panel) Amplitude of the nonadiabatic corrections in the present model (blue, solid line) and the one from Section~\ref{sec:Horava} (orange, dashed line) for $\lambda / \Lambda=0.01$, as a function of the {\it preferred} coordinate $X -X_K$, where $X_K$ denotes the position of the Killing horizon. In this example, the surface gravity is $\kappa \sim \Lambda$. We thus verify that the norm of the coefficient $\beta_\omo$ is of the order of the maximal value of the nonadiabatic parameter $\sim 10$, as expected from the analysis of nonadiabaticity~\cite{Massar:1997en}.
} \label{fig:characbis}
\end{figure}

To efficiently perform the numerical analysis, we regularized the metric and \ae ther field. In practice we worked with a metric of the form
\be 
ds^2 = \lp 1 - 2 f(r) \rp dv^2 - 2 dv dr,
\ee
and a unit norm \ae ther field
\be 
u^\mu \pd_\mu = \pd_v - f(r) \pd_r.
\ee
These expressions generalize the model of Section~\ref{sec:Horava} which is recovered for $f(r) =M/r$. The Killing horizon corresponds to $f(r)=1/2$, and the universal horizon to $f(r)=1$. We can then define the preferred coordinate $X$ along the lines of Appendix~\ref{app:PC}. For the numerical integration of \eq{eq:2DFE}, it is appropriate to work with $f$ expressed as a known function of $X$. A convenient choice is 
\be \label{eq:frX} 
f(r(X)) = \frac{1}{2} \lp 1- \eta \tanh \lp \frac{X}{X_0} \rp \rp.
\ee
$\eta$ is a positive parameter which must be equal to 1 to have an asymptotically flat space at $r \to \infty$ and a universal horizon at $r \to M$, i.e., $X \to -\infty$. In our numerical simulations, we worked with $\eta < 1$ to avoid large numerical errors due to the divergence of the dispersive roots $P^{\rm in}_{\pm \omo}$ close to the universal horizon. We then checked that the scattering coefficients become independent on $\eta$ in the limit $\eta \to 1$, as should be the case since the WKB approximations become exact on both sides. The advantage of this model is that the metric coefficients and the \ae ther field converge exponentially to their asymptotic values so that the asymptotic modes become exact solutions for $r \to \infty$, provided the decay rate of the exponentially decaying mode is small enough.~\footnote{Notice that an exponential convergence for $X \to -\infty$ is required to have a universal horizon at a finite value of $r = r_U$, such that $df/dr (r=r_U) \neq 0$.} The characteristics outside the universal horizon $r=r_U$ are shown in \fig{fig:characbis}. They exhibit the main properties of \fig{fig:char}. In the right panel we show the parameter governing nonadiabatic corrections, $\lp \pd_X P_\omo \rp / P_\omo^2$, as a function of the preferred coordinate $X$ in the present model and the one of Section~\ref{sec:Horava}. In the present model, they go to zero exponentially both for $X \to \infty$ and $X \to -\infty$. In the model of Section~\ref{sec:Horava}, they decay exponentially at $X \to -\infty$ but only polynomially at $X \to \infty$. The two models become equivalent, in the sense that the value of $P_\omo^u(r)$ follows 
the same law close to the Killing horizon when working with the same surface gravity.

\begin{figure}
\includegraphics[width=0.48 \linewidth]{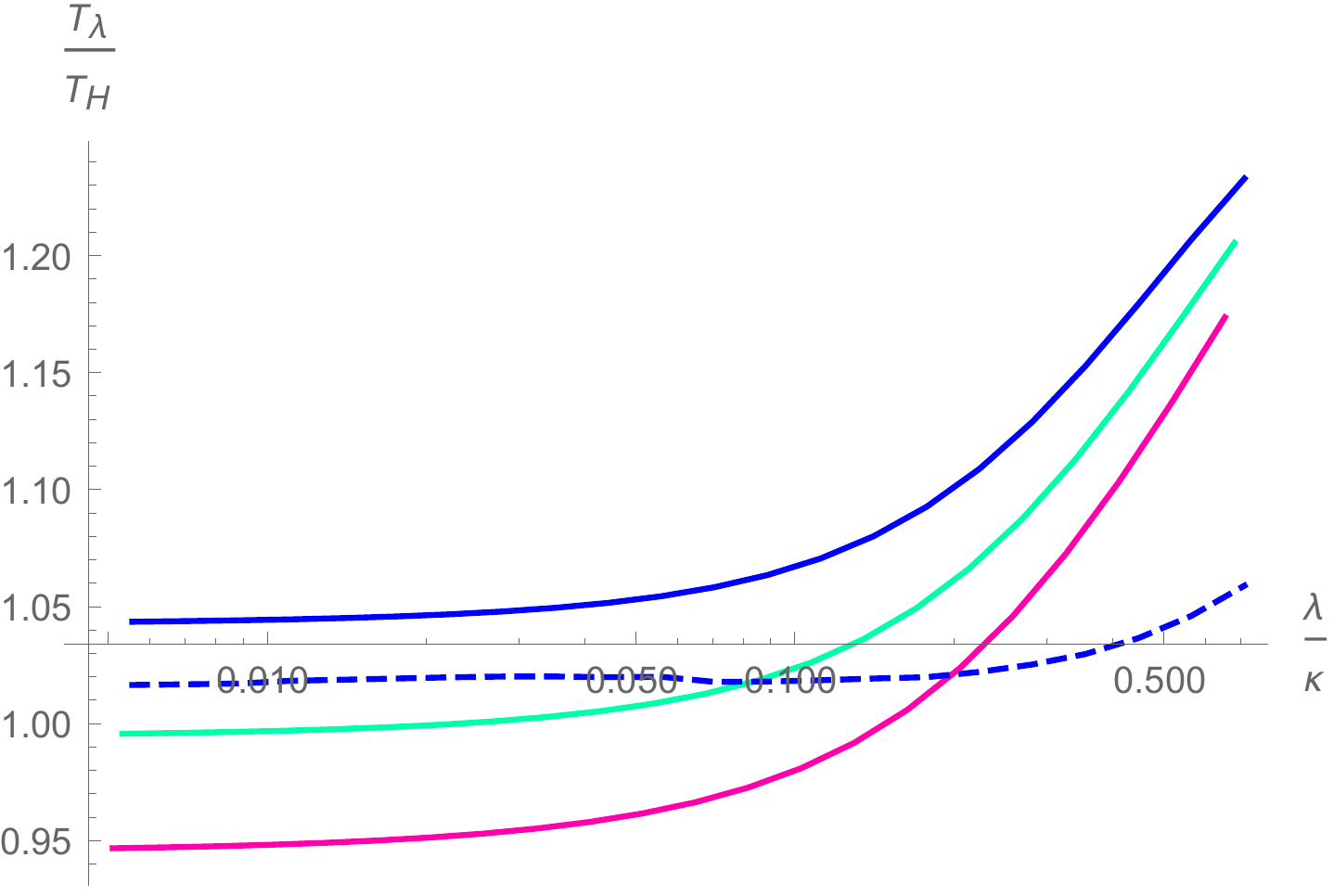} 
\caption{Plot of the effective temperature $T_\omo$ of \eq{eq:Teff} divided by the Hawking temperature as a function of $\lambda / \kappa$, where $\kappa$ is the surface gravity. The values of $\Lambda/\kappa$ are $1/2$ (solid line) and $3/2$ (dashed line). For the smallest value of $\Lambda$, we show the results for $\eta = 0.9$ (blue line), $0.94$ (cyan line), and $0.98$ (magenta line). For the largest value of $\Lambda$, these three curves are undistinguishable up to numerical errors. This indicates that the limit of the regulator $\eta \to 1^-$ is well defined, which was checked using a larger range of values for $\eta \in (0.8, 0.99)$. Moreover, when increasing the dispersive scale $\Lambda$, we see that $T_\omo$ closely agrees with the Hawking value $\kappa/2\pi$ for a larger domain of Killing frequencies.
} \label{fig:KillingScattering}
\end{figure}

The field equation was integrated numerically using~\cite{Mathematica10}, and the same techniques as in~\cite{Finazzi:2012iu,Michel:2014zsa}. The results are shown in \fig{fig:KillingScattering}. We obtain two important results. First, at fixed $\Lambda$ and $\kappa$, the effective temperature defined by
\be \label{eq:Teff} 
\left\lvert \beta_\lambda \right\rvert^2 = \frac{1}{e^{\lambda/T_\lambda} - 1}, 
\ee 
becomes independent of the regulator $\eta$ as $\eta \to 1^{-}$. Second at low frequencies $\omo \ll \Lambda$, we get a Planckian spectrum, i.e., $T_\lambda = constant$, with deviations from the Hawking temperature compatible with the results of~\cite{Macher:2009tw,Finazzi:2012iu}. This establishes that the propagation between the two horizons does not alter the thermal character of the outgoing spectrum.

\begin{figure}
\includegraphics[width=0.5 \linewidth]{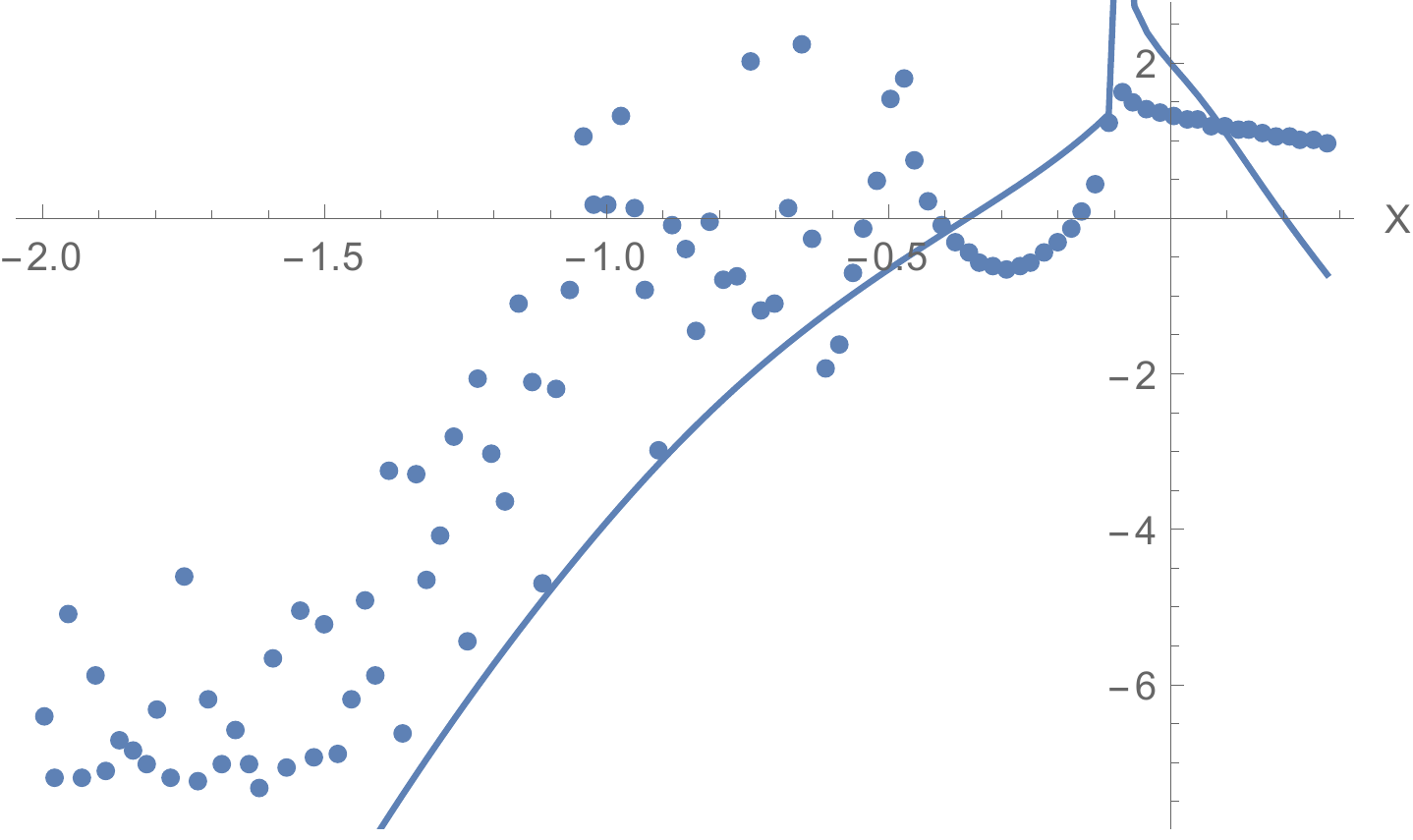} 
\caption{Deviations from the WKB approximation. The solid line shows the natural logarithm of $\left\lvert \lp \pd_X P_\omo^u \rp / \lp P_\omo^u \rp^2 \right\rvert$ as a function of the preferred coordinate $X$ in units of $1/\Lambda$, and the points show the logarithm of the relative difference between the solution computed numerically and the corresponding sum of the WKB modes, for $\Lambda=1$, $X_0 = 0.5$, $\eta = 0.99$, and  $\omo=0.1$. The important spread seems to be due to numerical errors a result of the increase of the momentum as $X \to -\infty$. The Killing horizon is located at $X=0$. 
} \label{devWKB}
\end{figure}

To conclude this numerical analysis, we numerically verify that the WKB approximation becomes exact when approaching the universal horizon. To this end, we show in \fig{devWKB} the logarithm of the relative deviation between the numerical solution and the weighted sum of the WKB waves $\alpha_\omo \psi^{u, \rm in}_{\omo} + \beta_\omo (\psi^{u, \rm in}_{-\omo})^*$. As $X \to -\infty$, we clearly see that the numerical values of the deviations decay following the rough estimation of the WKB corrections given by $\left\lvert \lp \pd_X P_\omo^u \rp / \lp P_\omo^u \rp^2 \right\rvert$. (The relatively important spread and the plateau for $X < -1.8$ seem to be due to numerical errors. Indeed, as the wave vector becomes very large, typically of order $10^2$, even a relatively small error in its value gives important and rapidly oscillating errors.) 

\bibliography{../biblio/bibliopubli}

\end{document}